\documentclass[twocolumn,pra,superscriptaddress,showpacs]{revtex4}
\usepackage{color}
\usepackage{amsmath}
\usepackage{amssymb}
\usepackage{graphicx}
\usepackage{esint}

\begin{document}

\title{Vacuum induced transparency and photon number resolved Autler-Townes
splitting in a three-level system}

\author{Jiang-Hao Ding}

\affiliation{Institute of Microelectronics, Tsinghua University, Beijing 100084,
China}

\affiliation{Institute of Applied Physics and Materials Engineering, FST, University
of Macau, Macau}

\author{Sai-nan Huai}

\affiliation{Institute of Microelectronics, Tsinghua University, Beijing 100084,
China}

\author{Hou Ian}

\affiliation{Institute of Applied Physics and Materials Engineering, FST, University
of Macau, Macau}

\author{Yu-xi Liu}
\email{yuxiliu@mail.tsinghua.edu.cn}

\affiliation{Institute of Microelectronics, Tsinghua University, Beijing 100084,
China}

\affiliation{Tsinghua National Laboratory for Information Science and Technology
(TNList), Beijing 100084, China}

\begin{abstract}
We study the absorption spectrum of a probe field by a $\Lambda$-type three-level system, which is coupled to
a quantized control field through the two upper energy levels.
The probe field is applied to the ground and the second excited states. When the quantized control field
is in vacuum, we derive a threshold condition to discern vacuum induced transparency (VIT) and vacuum induced
Autler-Townes splitting (ATS). We also find that the parameter change from VIT to vacuum induced ATS is
very similar to that from broken $PT$ symmetry to $PT$ symmetry. Moreover, we find the photon number resolved spectrum in the parameter regime of vacuum induced ATS when the mean photon number of the quantized control field is changed from zero (vacuum) to a finite number. However, there is no photon number resolved spectrum in the parameter regime of VIT even that the quantized control field contains the finite number of photons. Finally, we further discuss possible experimental realization.
\end{abstract}

\maketitle

\section{Introduction}

Electromagnetically induced transparency (EIT)~\cite{harris90,harris92,harris3,gong08}
is extensively studied in various three-level systems~\cite{Rieveiw1,Rieveiw2}, which interact with a classical control field and a probe field. If a medium can be modeled as an ensemble of many identical three-level atoms, then using EIT mechanism and under appropriate conditions, we can find that the absorption of the probe field by the medium can be
reduced, and the medium can becomes effectively transparent at the zero absorption
for the probe field, e.g., the probe field can freely pass through the medium. Thus, EIT is usually used to eliminate
the medium effect on the probe field by a classical control field.
If the absorption is monitored, then the reduction in the absorption is characterized by a dip~\cite{Abi}
in the absorption spectrum, and also there are two peaks surrounding the dip.

It is known that EIT is closely related to Autler-Townes splitting
(ATS)~\cite{ATS}. Both EIT and ATS study the absorption
of the probe field by a driven three-level system. Both of them display a reduction in absorption, and
have two peaks and a dip in their absorption spectra~\cite{Abi}.
However, the physical mechanisms
of their peaks and dip are different.
In ATS, the peaks and dip are due to two resonances corresponding to energy level splitting, induced
by the strong control field. However, the peaks and dip in EIT are due to the quantum
interference when two resonances in ATS become overlapped. The quantum interference can be understood via
dressed states~\cite{COHEN}, formed by the three-level system and control
field. That is, one resonant transition path, linked by the probe field in bare atomic
energy levels when the control field is not applied, becomes into two resonant transition paths in dressed state
picture when the control field is applied, and thus the resonant absorption of the probe field in bare atom might be
canceled through the quantum interference of transitions from these two paths. The threshold condition to distinguish EIT from ATS has been theoretically
explored~\cite{ATS-EIT,sun14} and also experimentally studied~\cite{Giner-2013,Sahin-Yang,Pei-2013,Qichunliu}.
ATS occurs when the strength of the control field is larger than the critical value.
EIT may appear when the strength of the control field is smaller than the critical value. Roughly speaking, the strength
of the control field for EIT is smaller than that for ATS.
A natural question is whether two peaks and a dip can
still appear when the photon number in the control field is finite or
is further reduced to zero? The answer to this question is related to the interaction
between the quantized control field and the three-level system.

We know that quantized fields can also form dressed states with atoms.
However, dressed states composed by quantized field and atom are very
different from those composed by classical field and atom. For example,
a dressed two-level system by a classical field still possesses
a character of two energy levels~\cite{COHEN,yxliu06}, which was experimentally
demonstrated~\cite{wilson07}, for example, via two-level superconducting
quantum circuits~\cite{you-today,clarke08,you-nature}. Whereas,
a dressed two-level system by a quantized field possesses many energy
levels. Using a structure of three energy levels, chosen from many
energy levels of dressed two-level systems by a quantized field~\cite{OC},
both EIT and ATS have been theoretically studied~\cite{ian10,H.Y.Zhang}.
Moreover, dressed two-level systems by quantized fields are also used
to generate stimulated amplification~\cite{oelsner13}, demonstrate
attenuation effects~\cite{shevchenko14}, and exhibit polariton states
with selective radiation spectrum~\cite{ian14}.

Let us come back three-level system with $\Lambda$ transitions for EIT and ATS. If the classical control field in such a system is replaced
by a quantized control field,  then the so-called vacuum
induced transparency (VIT) resulted from the quantum interference has been theoretically and experimentally
studied when the quantized control field
is in vacuum and the coupling strength between three-level system and quantized control field
is in the weak coupling regime~\cite{Vacuum-t,Vacuum-e}. However, to our knowledge,
the energy-level splitting induced by vacuum is not studied when such three-level system and quantized control field
is in the strong coupling regime. Hereafter, we call vacuum induced energy level splitting as vacuum induced ATS in analogy to ATS.
Furthermore, the threshold condition to discern VIT from vacuum induced ATS is not studied. Although
the photon number effect of the quantized control field on the quantum interference
is mentioned in Ref.~\cite{Vacuum-e}, and also the dependence of the EIT group delay on the photon
number is studied in Ref.~\cite{Nikoghosyan2010}, there is no study about the
photon number effect on the energy level splitting in analogy to ATS.

Here, we study absorption spectrum of a probe field by a $\Lambda$-type three-level
system, which is coupled to a quantized control field through the first and second excited states.
When the quantized control field is in vacuum, we first derive a threshold condition to discern
VIT from vacuum induced ATS, then we further analyze how VIT is changed to vacuum
induced ATS when the decay rate of the cavity field or the coupling
strength between the quantized control field and three-level system is varied.
This analysis is very similar to that for $PT$ symmetric systems,
e.g., in Refs.~\cite{Cal,lan,jiang}. When the quantized control field contains finite number of photons,
we first show how the photon number affects the absorption spectrum when the coupling strength is in the
VIT parameter regime. In particular,
we will show that the photon number resolved spectrum can be found
when the coupling strength is in the parameter regime of vacuum induced ATS.
In view of experimental progresses of atomic physics and
superconducting quantum devices~\cite{you-today,clarke08,you-nature}
and circuit quantum electrodynamics (CQED)~\cite{circuitQED-R},
on which ATS~\cite{ATS-1,ATS-2,ATS-3,ATS-4,ATS-5,ATS-6,ATS-7}, population
trap~\cite{ATS-8}, adiabatic population transfer~\cite{Zhao2016},and EIT~\cite{Xiugu,SEIT} have been theoretically
and experimentally demonstrated, we will also discuss possible experimental
realization of VIT and vacuum induced ATS in either atomic systems or superconducting CQED
systems.

Our paper is organized as follows. In Sec.~II, we
describe theoretical model and write out dressed states of the studied
system. In Sec.~III, we give a definition of the susceptibility,
and then show detailed steps for deriving a formula to describe the susceptibility. In Sec.~IV,
we analyze the properties of the absorption spectrum when the quantized control field is in vacuum. Similar to the asymmetric profile of the
absorption spectrum~\cite{Knight,minxiao} for classically driven
three-level system, we give a detailed analysis on the asymmetric
profile of two resonances. Then, we derive a threshold condition to
discern VIT from vacuum induced ATS. In Sec.~V, we analyze photon number effect on EIT and ATS by
incoherently or coherently pumping the quantized control field. In
particular, we show the photon number resolved spectrum when the coupling strength between quantized control field
and three-level system is in the parameter regime of vacuum induced ATS. We will also analyze the reason why there is
no photon number resolved spectrum when the coupling strength between the quantized controlled field and three-level system is in the
parameter regime of VIT. In Sec.~VI, we will apply our study to atomic systems and superconducting
CQED systems and discuss possible experimental realization. Finally,
we summarize the results in Sec.~VII.

\section{Hamiltonian of the system}

\begin{figure}
\includegraphics[width=7cm]{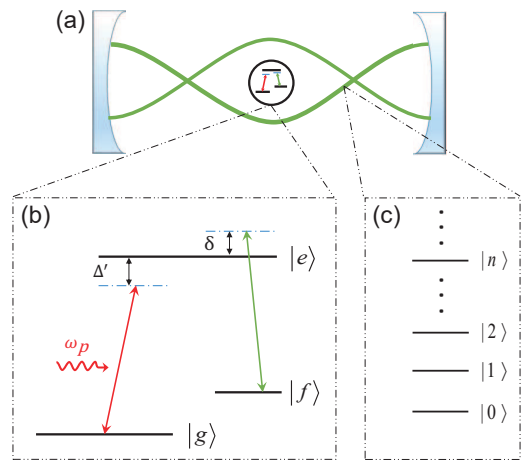}
\caption{(Color online) (a) A schematic diagram for a three-level system with
$\Lambda$-type transitions inside the cavity. Here two green curves
schematically represent the cavity field. (b) The schematic diagram
for three-level system coupled to a single-mode cavity field and a
classical probe field. The cavity field induces the transition between
the energy levels $|e\rangle$ and $|f\rangle$, however the probe
field induces the transition between the energy levels $|e\rangle$
and $|g\rangle$. Here, $\Delta^{\prime}=\omega_{e}-\omega_{p}$ is
the detuning between the probe field with the frequency $\omega_{p}$
and the transition frequency $\omega_{e}$ of the three-level system,
$\delta=\omega_{c}-(\omega_{e}-\omega_{f})$ denotes the detuning
between the cavity field with the frequency $\omega_{c}$ and the
transition frequency $\omega_{e}-\omega_{f}$ of the three-level system.
(c) A schematic diagram for the energy levels of the cavity field
with the equal level spacing. }
\label{fig1}
\end{figure}

As schematically shown in Fig.~\ref{fig1}, we study a $\Lambda$-type
three-level system, which is placed inside a cavity. The ground state,
first and second excited states of the three-level system are denoted
by $|g\rangle$, $|f\rangle$ and $|e\rangle$. For generality of
the study, we first do not specify this three-level system to a particular
physical object. We further assume that the quantized single-mode
cavity field with frequency $\omega_{c}$ induces the transition from
the energy level $\left\vert f\right\rangle $ to the energy level
$\left\vert e\right\rangle $, while a weak classical probe field
with frequency $\omega_{p}$ induces the transition between the energy
levels $\left\vert g\right\rangle $ and $\left\vert e\right\rangle $.
Later on, we call the quantized single-mode cavity field as the quantized control field or
the cavity field.
Under the rotating-wave approximation, the Hamiltonian of the whole
system is given by
\begin{equation}
H=H_{c}+H_{p},\label{ham_twopart}
\end{equation}
with
\begin{eqnarray}
H_{c} & = & \hbar\omega_{c}a^{\dagger}a+\hbar\omega_{e}\left\vert e\right\rangle \left\langle e\right\vert +\hbar\omega_{f}\left\vert f\right\rangle \left\langle f\right\vert \label{ham_cav}\\
 &  & +\hbar\eta(\left\vert e\right\rangle \left\langle f\right\vert a+\text{h.c.}),\nonumber
\end{eqnarray}
and
\begin{equation}
H_{p}=\hbar\varepsilon(\left\vert e\right\rangle \left\langle g\right\vert e^{-i\omega_{p}t}+\text{h.c.}),\label{ham_probe}
\end{equation}
where $\omega_{f}$ and $\omega_{e}$ are the transition frequencies
from the first excited state $\left\vert f\right\rangle $ and the
second excited state $\left\vert e\right\rangle $ to the ground state
$|g\rangle$, respectively. The parameter $\eta$ denotes the coupling
strength between the three-level system and the cavity field, $\varepsilon$
denotes the coupling strength between the probe field and the three-level
system.

\begin{figure}
\includegraphics[width=8.5cm]{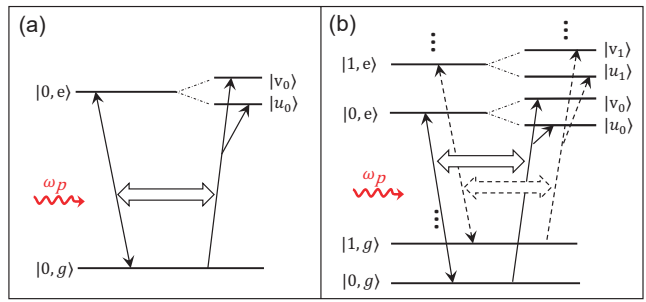}
\caption{Schematic diagrams for the coupling between the dressed three-level
system and the weak probe field. (a) shows that
the probe field, used to induce the transition between the ground
state $\left\vert g\right\rangle $ and the excited state $\left\vert e\right\rangle $
when the cavity field does not exist, is changed to induce the transitions
from the ground state $\left\vert 0,g\right\rangle $ to the states
$\left\vert u_{0}\right\rangle $ and $\left\vert v_{0}\right\rangle $, respectively,
in the dressed state basis when the cavity field is in vacuum. (b) shows that
the transition from the state $\left\vert n,g\right\rangle $ to
the state $\left\vert n,e\right\rangle $ are changed to the transitions from the
state $\left\vert n,g\right\rangle $ to the states $\left\vert u_{n}\right\rangle $
and $\left\vert v_{n}\right\rangle $, respectively, in the dressed state basis when the
cavity field contains the finite number of photons. Here, as an example, we just take $n=0$
and $n=1$.}
\label{fig2}
\end{figure}

It is obvious that the probe field can be resonantly absorbed by the
three-level system when the cavity field is not coupled to the three-level
system, i.e., $\eta=0$. To clearly show the effect of the cavity
field on the absorption of the probe field, we now rewrite the Hamiltonian
in Eq.~\eqref{ham_cav} in the dressed state basis, formed by the
cavity field and two upper energy levels $|e\rangle$ and $|f\rangle$
of the three-level system. That is, the Hamiltonian in Eq.~(\ref{ham_cav})
can be rewritten as
\begin{equation}
H_{c}=\sum\limits _{n}(E_{-,n}\left\vert u_{n}\right\rangle \left\langle u_{n}\right\vert +E_{+,n}\left\vert v_{n}\right\rangle \left\langle v_{n}\right\vert ),\label{eq:4}
\end{equation}
in the dressed state basis $|u_{n}\rangle$ and $|v_{n}\rangle$,
given by
\begin{eqnarray}
\left\vert u_{n}\right\rangle  & = & \cos\theta_{n}\left\vert n,e\right\rangle -\sin\theta_{n}\left\vert n+1,f\right\rangle ,\label{eq:un}\\
\left\vert v_{n}\right\rangle  & = & \sin\theta_{n}\left\vert n,e\right\rangle +\cos\theta_{n}\left\vert n+1,f\right\rangle ,\label{eq:vn}
\end{eqnarray}
with $n$-dependent parameter
\begin{equation}
\theta_{n}=\frac{1}{2}\tan^{-1}\left[\frac{2\eta\sqrt{n+1}}{\delta}\right],\label{eq:trans_angle}
\end{equation}
and $\delta=\omega_{c}-(\omega_{e}-\omega_{f})$. Here, the state
$|n+1,f\rangle$ (or $|n,e\rangle$) denotes that there are $n+1$
(or $n$) photons inside the cavity and the three-level system is
in the state $|f\rangle$ (or $|e\rangle$). Later on, for convenience,
we call $|u_{n}\rangle$ and $|v_{n}\rangle$ as $n$-photon dressed
states. The eigenvalues $E_{-,n}$ and $E_{+,n}$, corresponding to
eigenstates $|u_{n}\rangle$ and $|v_{n}\rangle$, are given by
\begin{eqnarray}
E_{\pm,n} & = & (n+\frac{1}{2})\hbar\omega_{c}+\frac{\hbar}{2}(\omega_{e}+\omega_{f})\nonumber \\
 &  & \pm\frac{\hbar}{2}\sqrt{\delta^{2}+4\eta^{2}(n+1)}.\label{eq:eigenvalues}
\end{eqnarray}
Using dressed states in Eq.~\eqref{eq:un} and Eq.~\eqref{eq:vn} as basis,
the Hamiltonian $H_{p}$ in Eq.~(\ref{ham_probe}) can be rewritten
as
\begin{equation}
H_{p}=\hbar\varepsilon\sum\limits _{n}\left[(\cos\theta_{n}\left\vert u_{n}\right\rangle +\sin\theta_{n}\left\vert v_{n}\right\rangle )\left\langle n,g\right\vert e^{-i\omega_{p}t}+\text{h.c.}\right],\label{eq:hp_dressed}
\end{equation}
by replacing $|n,e\rangle$ with $\left\vert u_{n}\right\rangle $
and $\left\vert v_{n}\right\rangle $. We note that $\left\vert n,g\right\rangle $
is not a dressed state, but it is orthogonal to $\left\vert u_{n}\right\rangle $
and $\left\vert v_{n}\right\rangle $. This is because $\left\vert u_{n}\right\rangle $
and $\left\vert v_{n}\right\rangle $ are linear superposition of
$\left\vert n,e\right\rangle $ and $\left\vert n+1,f\right\rangle $,
which are orthogonal to $\left\vert n,g\right\rangle $.
Here, we will study the absorption spectrum of the whole system to the
probe field when the cavity field is in vacuum or contains the finite number of
photons, as schematically shown in Fig.~\ref{fig2}. The basic mechanism of absorption
for two cases can be qualitatively explained as below.

If  there is no photon inside the cavity or the cavity field is in vacuum,
then all populations are in the ground state $\left\vert 0,g\right\rangle $
when the whole system reaches the steady state. Thus, the probe field,
which induces the transition between the states $\left\vert g\right\rangle $
and $\left\vert e\right\rangle $ when there is no cavity field, is changed to link the transition from
the state $\left\vert 0,g\right\rangle $ to the state $\left\vert 0,e\right\rangle $ when the cavity field
is coupled. Thus, in the dressed state basis, as schematically shown in Fig.~\ref{fig2}(a),
one transition path from the state $\left\vert 0,g\right\rangle $
to the state $\left\vert 0,e\right\rangle $ can be changed to two transition
paths from the state $\left\vert 0,g\right\rangle $ to the states
$\left\vert u_{0}\right\rangle $ and $\left\vert v_{0}\right\rangle $,
respectively. This is because $\left\vert 0,e\right\rangle $ can be expressed as the
superposition of $\left\vert u_{0}\right\rangle $ and $\left\vert v_{0}\right\rangle$.
In different parameter regimes of the coupling strength between the quantized control field and the
three-level system, these transitions might result in either VIT spectrum~\cite{Vacuum-t,Vacuum-e} or vacuum induced ATS spectrum.
In the following, we will give a detailed study for a threshold condition to discern them.

If there is the finite number $N$ of photons inside the cavity, all states $\left\vert n,g\right\rangle $,
$\left\vert n,f\right\rangle $ and $\left\vert n,e\right\rangle $
with $n=0,1,\cdots, N$ might be occupied with the certain probability
when the whole system reaches the steady state. Thus, the probe field,
which induces the transitions between the energy levels $|g\rangle$
and $|e\rangle$ when there is no cavity field, links $N$ transitions between the energy levels $|n,g\rangle$
and $|n,e\rangle$ ($n=0,1,\cdots, N$) when the cavity field is coupled. In the dressed state basis, as schematically shown in Fig.~\ref{fig2}(b),
each probe-field-induced transition from the state $|n,g\rangle$
to the state $|n,e\rangle$ is changed to transitions from the state
$|n,g\rangle$ to the dressed states $\left\vert u_{n}\right\rangle $
and $\left\vert v_{n}\right\rangle $, respectively. In this case, the absorption
spectrum of the probe field should be sum of $2N$ transitions from
the state $|n,g\rangle$ to the dressed states $\left\vert u_{n}\right\rangle $
and $\left\vert v_{n}\right\rangle $ with $n=0,1,\cdots,N$. These $2N$ transitions
play different roles in the parameter regime of VIT and vacuum induced ATS. The quantitative analysis will
be given below.

\section{Susceptibility and master equation}

The key parameter to characterize the absorption of a probe field by a system
is the linear susceptibility $\chi$. Its real
and imaginary parts represent the dispersion and the absorption
of the probe field, respectively. When the probe field is applied
to a three-level system via the transition from the state $|g\rangle$
to the state $|e\rangle$, the linear susceptibility $\chi$ of the three-level
system is given as
\begin{equation}
\chi=\frac{\mu_{ge}}{\epsilon_{0}\varepsilon}\rho_{ge},\label{eq:susceptibility}
\end{equation}
where $\mu_{ge}$, $\epsilon_{0}$, and $\varepsilon$ are the dipole
moment from the state $|g\rangle$ to the state $|e\rangle$ of the three-level system, the vacuum permittivity, and the
coupling strength between the probe field and the three-level system.
Thus, $\chi$ is determined by $\rho_{ge}$. When the
quantized control field is coupled to the tree-level system, the matrix element $\rho_{ge}$ in Eq.~\eqref{eq:susceptibility}
can be further expressed as
\begin{equation}
\rho_{ge}=\sum_{n=0}^{\infty}\left\langle n,g\right\vert \rho\left\vert e,n\right\rangle .\label{eq:rho_ge_npho}
\end{equation}
We first show how to obtain $\rho_{eg}$ by solving the master equation
\begin{eqnarray}
\dot{\rho} & = & \frac{1}{i\hbar}[H,\rho]+\kappa(2a\rho a^{\dagger}-a^{\dagger}a\rho-\rho a^{\dagger}a),\label{eq:master_equation}\\
 &  & +\sum\limits _{i,j=e,f,g}^{i\geqslant j}\frac{\gamma_{ij}}{2}[2\sigma_{ji}\rho\sigma_{ij}-\sigma_{ij}\sigma_{ji}\rho-\rho\sigma_{ij}\sigma_{ji}],\nonumber
\end{eqnarray}
for the reduced density matrix $\rho$ of the system in the Born-Markov
approximation when the system is at zero temperature. We notice that incoherent pumping (the finite temperature effect)
or coherent pumping to the quantized control field will be studied in later of this manuscript. Here, the Hamiltonian
$H$ is given by Eq.~(\ref{ham_twopart}). In Eq.~\eqref{eq:master_equation},
$\sigma_{ij}=\left\vert i\right\rangle \left\langle j\right\vert $
is the ladder operator of the three-level system, where $\left\vert i\right\rangle $
and $\left\vert j\right\rangle $ are one of the states $\left\vert g\right\rangle $,
$\left\vert f\right\rangle $, and $\left\vert e\right\rangle$ with the order from the
ground to the second excited state;
$\gamma_{ij}$ denotes the decay rate of the three-level system when
there is no cavity field. For example, $\gamma_{eg}$ denotes the
decay rate from the state $|e\rangle$ to the state $|g\rangle$ when
there is no cavity field. $\kappa$ represents the decay rate of the
quantized single-mode cavity field.

As discussed in Sec.~II, the absorption spectrum of the probe field by quantized-field-controlled three-level system can be
more conveniently solved in dressed state basis. Thus, using the states $|n,g\rangle$, the dressed states $|u_{n}\rangle$ and $|v_{n}\rangle$ in Eq.~\eqref{eq:un} and
Eq.~\eqref{eq:vn}, formed by states $\left\vert n,e\right\rangle ,\left\vert n+1,f\right\rangle $ of the whole system as the basis,  the operators $\sigma_{ij}$, $a^{\dagger}$ and their hermitian conjugations in the master equation~(\ref{eq:master_equation})
can be rewritten via the relations
\begin{eqnarray}
\sigma_{ij} & = & \left\vert i\right\rangle \left\langle j\right\vert =\sum_{n}\left\vert n,i\right\rangle \left\langle j,n\right\vert ,\label{eq:sigma_ij}\\
a^{\dagger} & = & \sum_{n,i}\sqrt{n+1}\left\vert n+1,i\right\rangle \left\langle i,n\right\vert .\label{eq:adagger}
\end{eqnarray}
We note that the completeness conditions $\sum_{n}|n\rangle\langle n|=1$
and $\sum_{i}|i\rangle\langle i|=1$ for the states $|n\rangle$ and
$|i\rangle$ of the single-mode cavity field and three-level system
are used when Eqs.~(\ref{eq:sigma_ij}) and (\ref{eq:adagger}) are
derived. Substituting expressions of the Hamiltonian $H$, $a^{\dagger}$,
$\sigma_{ij}$ and their hermitian conjugations in dressed state basis
into Eq.~\eqref{eq:master_equation}, we can have equations of motion
for the matrix elements in dressed state basis.

At the zero temperature as shown in Eq.~(\ref{eq:master_equation}), the whole system is in vacuum
and only the ground $|0,g\rangle$ is populated when the whole system
is in the steady state. As discussed in Fig.~\ref{fig2}(a) of Sec.~II,
the subspace to study absorption spectrum of the probe field is limited to the
subspace formed by three basis states $\left\vert 0,g\right\rangle ,\left\vert 0,e\right\rangle ,\left\vert 1,f\right\rangle $,
which can be rewritten in terms of $\left\vert 0,g\right\rangle \equiv\left\vert G\right\rangle $,
$\left\vert u_{0}\right\rangle =\cos\theta_{0}|0,e\rangle-\sin\theta_{0}|1,f\rangle\equiv\left\vert u\right\rangle $
and $\left\vert v_{0}\right\rangle =\sin\theta_{0}|0,e\rangle+\cos\theta_{0}|1,f\rangle\equiv\left\vert v\right\rangle $
in the dressed state basis. Thus the equations of motion for matrix elements
$\rho_{Gv}$ and $\rho_{Gu}$ are given by
(the detailed derivations are shown in Appendix A)
\begin{eqnarray}
\dot{\rho}_{Gv} & = & \frac{1}{i\hbar}[-E_{+,0}\rho_{Gv}+(\rho_{vv}-\rho_{GG})\hbar\varepsilon\sin\theta_{0}e^{i\omega_{p}t}\nonumber \\
 &  & +\hbar\varepsilon\cos\theta_{0}e^{i\omega_{p}t}\rho_{uv}]\nonumber \\
 &  & +\Gamma_{\mathrm{c}}\rho_{Gu}+\Gamma_{Gv}\rho_{Gv},\label{eq:rho_gv}
\end{eqnarray}
\begin{eqnarray}
\dot{\rho}_{Gu} & = & \frac{1}{i\hbar}[-E_{-,0}\rho_{Gu}+(\rho_{uu}-\rho_{GG})\hbar\varepsilon\cos\theta_{0}e^{i\omega_{p}t}\nonumber \\
 &  & +\hbar\varepsilon\sin\theta_{0}e^{i\omega_{p}t}\rho_{vu}]\nonumber \\
 &  & +\Gamma_{\mathrm{c}}\rho_{Gv}+\Gamma_{Gu}\rho_{Gu}.\label{eq:rho_gu}
\end{eqnarray}
Here, we define the relaxation rates $\Gamma_{Gv}$ and $\Gamma_{Gu}$
as
\begin{eqnarray*}
\Gamma_{Gv} & = & -\gamma_{e}\sin^{2}\theta_{0}-(\gamma_{f}+\kappa)\cos^{2}\theta_{0},\\
\Gamma_{Gu} & = & -\gamma_{e}\cos^{2}\theta_{0}-(\gamma_{f}+\kappa)\sin^{2}\theta_{0},
\end{eqnarray*}
and the relaxation rate
\[
\Gamma_{\mathrm{c}}=(-\gamma_{e}+\gamma_{f}+\kappa)\sin\theta_{0}\cos\theta_{0}.
\]
with $\gamma_{e}=(\gamma_{eg}+\gamma_{ef}+\gamma_{ee})/2$ and $\gamma_{f}=(\gamma_{fg}+\gamma_{ff})/2$.
Here, we assume $\gamma_{gg}=0$.

By solving Eqs.~(\ref{eq:rho_gv}) and (\ref{eq:rho_gu}) via perturbation
theory, we can obtain the density matrix $\rho_{ge}$ in vacuum case,
and then obtain the susceptibility $\chi$ for discussing VIT and
vacuum induced ATS by virtue of the imaginary
part ${\rm Im}[\chi]$.

\section{Vacuum induced transparency and Autler-Townes splitting}

\subsection{Symmetric or asymmetric absorption}

At the zero temperature, the quantized control field is in the vacuum and the occupation is only in the ground state $|0,g\rangle$ when the whole
system reaches the steady state. In this case, the susceptibility $\chi$ is proportional to the matrix
element
\begin{equation}
\rho_{ge}=\left\langle 0,g\right\vert \rho\left\vert e,0\right\rangle =\cos\theta_{0}\rho_{Gu}+\sin\theta_{0}\rho_{Gv},\label{eq:rho_ge}
\end{equation}
which is expressed in the zero-photon dressed state basis. When Eq.~(\ref{eq:rho_ge})
is derived, we express $|0,e\rangle$ as the superposition of zero-photon
dressed states $|u_{0}\rangle\equiv|u\rangle$ and $|v_{0}\rangle\equiv|v\rangle$, that is,
$|0,e\rangle=\cos\theta_{0}|u\rangle+\sin\theta_{0}|v\rangle$.

It is obvious that $\rho_{ge}$ can be straightforwardly
obtained by solutions of $\rho_{Gu}$ and $\rho_{Gv}$, which can
be given by solving Eqs.~(\ref{eq:rho_gv}) and (\ref{eq:rho_gu})
using perturbation theory for different orders of the strength $\varepsilon$
of the probe field, i.e., $\rho_{ij}=\sum_{m=0}\varepsilon^{m}\rho_{ij}^{(m)}$
with the subscript $ij=Gv$ or $ij=Gu$. The zero-order solution $\rho_{ij}^{(0)}$is
a steady state solution of the system when the probe field is not
applied to the system. $\rho_{ij}^{(0)}$ can be obtained by solving
Eqs.~(\ref{eq:rho_gv}) and (\ref{eq:rho_gu}) with assumption $\partial\rho_{ij}/\partial t=0$
and $\varepsilon=0$. Using the dressed state basis, we obtain $\rho_{GG}^{(0)}\approx1$
and $\rho_{uu}^{(0)}=\rho_{vv}^{(0)}=\rho_{uv}^{(0)}=0$. Substituting
these values of the matrix elements into Eqs.~(\ref{eq:rho_gv})
and ~(\ref{eq:rho_gu}), and then solving the equations of motion
up to the first order of $\varepsilon$, we have
\begin{equation}
\rho_{Gv}=\frac{i\varepsilon\{\sin\theta_{0}[i(\omega_{-,0}-\omega_{p})+\Gamma_{Gu}]-\cos\theta_{0}\Gamma_{\mathrm{c}}\}}{\Gamma_{\mathrm{c}}^{2}-[i(\omega_{-,0}-\omega_{p})+\Gamma_{Gu}][i(\omega_{+,0}-\omega_{p})+\Gamma_{Gv}]},\label{eq:rho_gv_solution}
\end{equation}
\begin{equation}
\rho_{Gu}=\frac{i\varepsilon\{\cos\theta_{0}[i(\omega_{+,0}-\omega_{p})+\Gamma_{Gv}]-\sin\theta_{0}\Gamma_{\mathrm{c}}\}}{\Gamma_{\mathrm{c}}^{2}-[i(\omega_{-,0}-\omega_{p})+\Gamma_{Gu}][i(\omega_{+,0}-\omega_{p})+\Gamma_{Gv}]},\label{eq:rho_gu_solution}
\end{equation}
where $\omega_{\pm,0}=E_{\pm,0}/\hbar$. Combining Eqs.~\eqref{eq:rho_gv_solution}-\eqref{eq:rho_gu_solution}
with Eq.~\eqref{eq:rho_ge}, we obtain the analytic solution of $\rho_{ge}$
as
\begin{equation}
\rho_{ge}=\frac{\varepsilon[\Delta+i(\gamma_{f}+\kappa)+\delta/2]}{-\Delta^{2}-i(\gamma_{e}+\gamma_{f}+\kappa)\Delta+C},\label{eq:rho_ge_solution}
\end{equation}
with the detuning
\begin{equation}
\Delta=\frac{1}{2}(\omega_{+,0}+\omega_{-,0})-\omega_{p}.
\end{equation}
The parameter $C$ is given by
\begin{equation}
C=\eta^{2}+\gamma_{e}\gamma_{f}+\gamma_{e}\kappa+\frac{\delta^{2}}{4}+i\frac{\delta}{2}(-\gamma_{e}+\gamma_{f}+\kappa).\label{eq:22}
\end{equation}
We use Eq.~\eqref{eq:rho_ge_solution} and Eq.~\eqref{eq:susceptibility}
to obtain the imaginary part $\mathrm{Im}[\chi]$ of the susceptibility
$\chi$ as
\begin{eqnarray}
\mathrm{Im}[\chi]=Z\left[\gamma_{e}\left(\Delta+\frac{\delta}{2}\right)^{2}+\eta^{2}(\gamma_{f}+\kappa)+\gamma_{e}(\gamma_{f}+\kappa)^{2}\right],\label{eq:imag_chi}
\end{eqnarray}
which has equivalent form to the imaginary part of the susceptibility in Ref.~\cite{Vacuum-e}. Here
\begin{eqnarray*}
Z & = & \frac{\mu_{ge}}{\epsilon_{0}}\left\{ \left(-\Delta^{2}+\eta^{2}+\gamma_{e}\gamma_{f}+\frac{\delta^{2}}{4}+\gamma_{e}\kappa\right)^{2}\right.\\
 &  & +\left.\left[-\Delta(\gamma_{e}+\gamma_{f}+\kappa)+\frac{1}{2}\delta(-\gamma_{e}+\gamma_{f}+\kappa)\right]^{2}\right\} ^{-1}.
\end{eqnarray*}

\begin{figure}
\includegraphics[width=8cm]{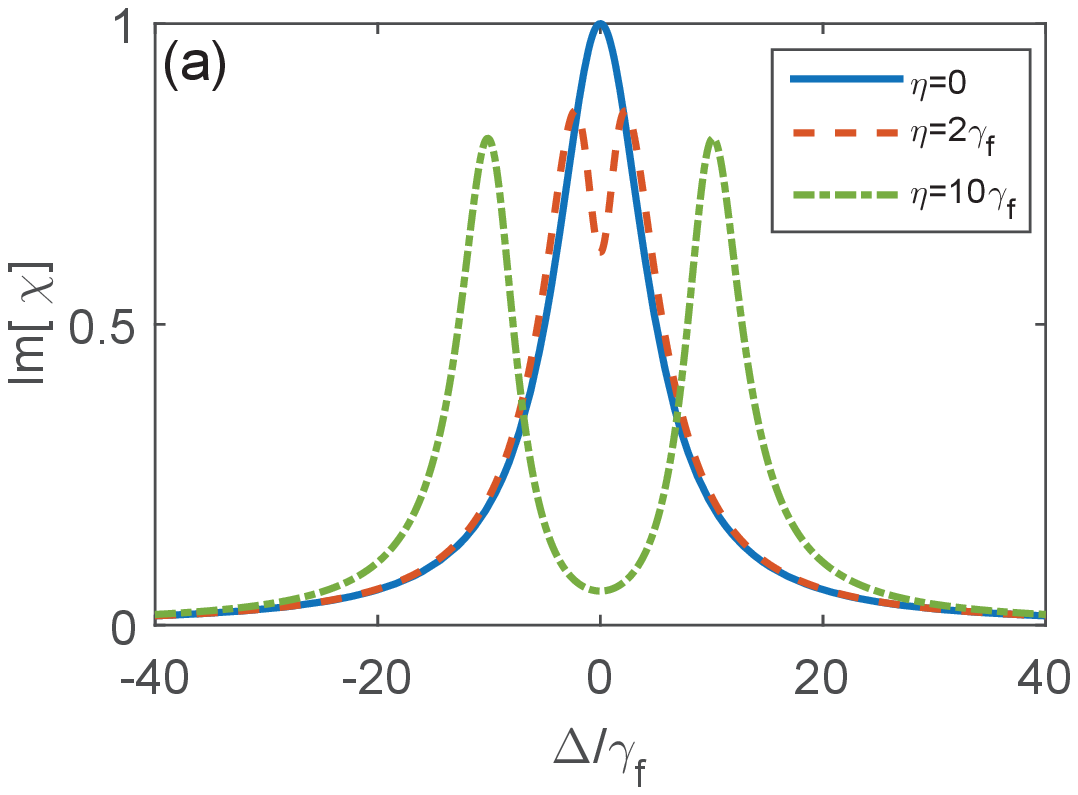} \includegraphics[width=8cm]{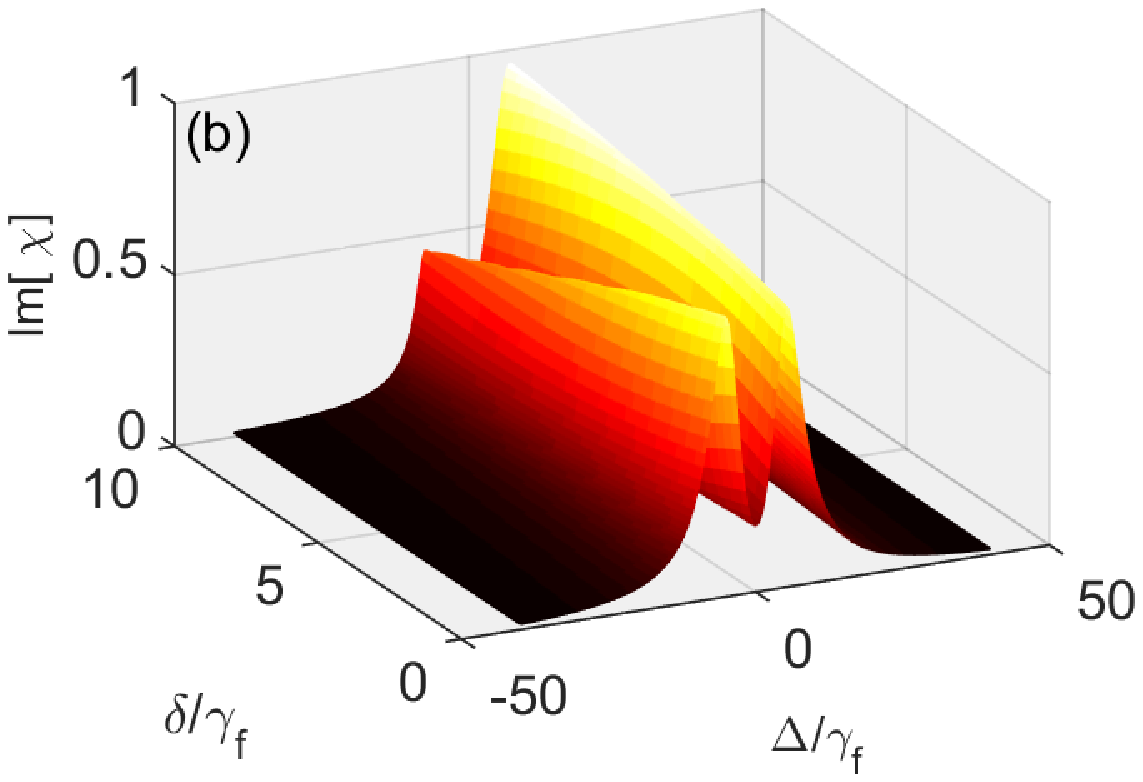}
\caption{(Color online) Illustration of the normalized absorption to the probe
field in vacuum case with unit $\gamma_{f}$. (a) shows the imaginary
part of the susceptibility $\chi$ versus the atom-probe detuning
$\Delta$ when $\delta=0$. The solid, dashed and dash-dotted curves
represent different coupling strengths $\eta=0$, $\eta=2\gamma_{f}$
and $\eta=10\gamma_{f}$ between the three-level system and the single-mode
cavity field. (b) shows the imaginary part of the susceptibility $\chi$
versus the detunings $\Delta$ ($\delta$) between the probe field
(the cavity field) and the three-level system. Here we assume $\gamma_{e}=5\gamma_{f}$
and $\kappa=0.2\gamma_{f}$.}
\label{fig3}
\end{figure}

In Fig.~\ref{fig3}(a), we show the variation of $\mathrm{Im}[\chi]$
with the detuning $\Delta$ when the cavity field resonantly interacts with
the three-level system and other parameters are given. The solid,
dashed and dash-dotted curves show the absorption spectra for different
coupling strengths $\text{\ensuremath{\eta}}$. If the three-level
system and the cavity field are decoupled (i.e. $\eta=0$), as shown
in the solid curve in Fig.~\ref{fig3}(a), the absorption profile
exhibits single absorption peak. The linewidth of the peak is roughly
proportional to $\gamma_{e}$. Figure~\ref{fig3}(a) also shows that
the absorption spectra are symmetric when the cavity field resonantly
interacts with the three-level system. With the increase of the coupling
strength $\eta$, the absorption profile begins to exhibit a dip.
For example, when $\eta=2\gamma_{f}$, there are two peaks and a sharp
dip in the center of the absorption spectrum, but the distance between two
peaks is smaller than $2\eta$. As $\eta$ is further increased to
$10\gamma_{f}$, the distance between two peaks becomes into $2\eta$.

In Fig.~\ref{fig3}(b), $\mathrm{Im}[\chi]$ is further plotted as
functions of the detunings $\Delta$ and $\delta$. We find that the
absorption spectrum is asymmetric when the cavity field does not resonantly
interact with the three-level system, that is, the asymmetric absorption
is due to the detuning between the three-level system and the single-mode
cavity field. As shown in Fig.~\ref{fig3}(b), the two peaks in the
absorption spectrum have the same height when $\delta=0$. With the
increase of $\delta$, the height of one peak is increased while the
other one is decreased, and then presents the asymmetric profile. We
note that the two absorption peaks for $\delta<0$ behave oppositely
in contrast to those for $\delta>0$. This asymmetric property is
very similar to that for a classically driven three-level system~\cite{Knight}
with the ladder-type transition when the driving field does not resonant
with the three-level system. By analyzing the imaginary part of the
susceptibility $\chi$, we find that when $\delta=0$, the expression
of $\mathrm{Im}[\chi]$ is an even function of $\Delta$, and has the symmetric
resonances. When $\delta\neq0$, the $\Delta\delta$ term in numerator
makes the expression of $\mathrm{Im}[\chi]$ neither an even function
nor an odd function of $\Delta$. This results in profile.

We mention that the spectrum in Ref.~\cite{Knight} is calculated
in the rotating reference frame with the frequency of the control
field, the two asymmetric resonances have the same distance relevant
to the shifted origin (the origin of the rotating reference frame)
when the control field is unresonant with the three-level system,
thus two asymmetric resonances have different distances relevant to
unshifted origin. This results in an observation that two asymmetric
resonances in Ref.~\cite{Knight} are shifted by an unequal amount
from the unperturbed resonance at zero probe-detuning. However, in
our calculation, we work in the laboratory picture, the two asymmetric
resonances have the same distance relevant to the unshifted origin,
that is, they have an equal amount frequency shift from the unperturbed
resonance at zero probe-detuning.

Below, we will further analyze the reason why there are different
distances between two peaks of the absorption spectra in different coupling
strengths, as shown in Fig.~\ref{fig3}(a). These are related to
VIT and vacuum induced ATS.

\begin{figure*}
\includegraphics[width=6.5cm]{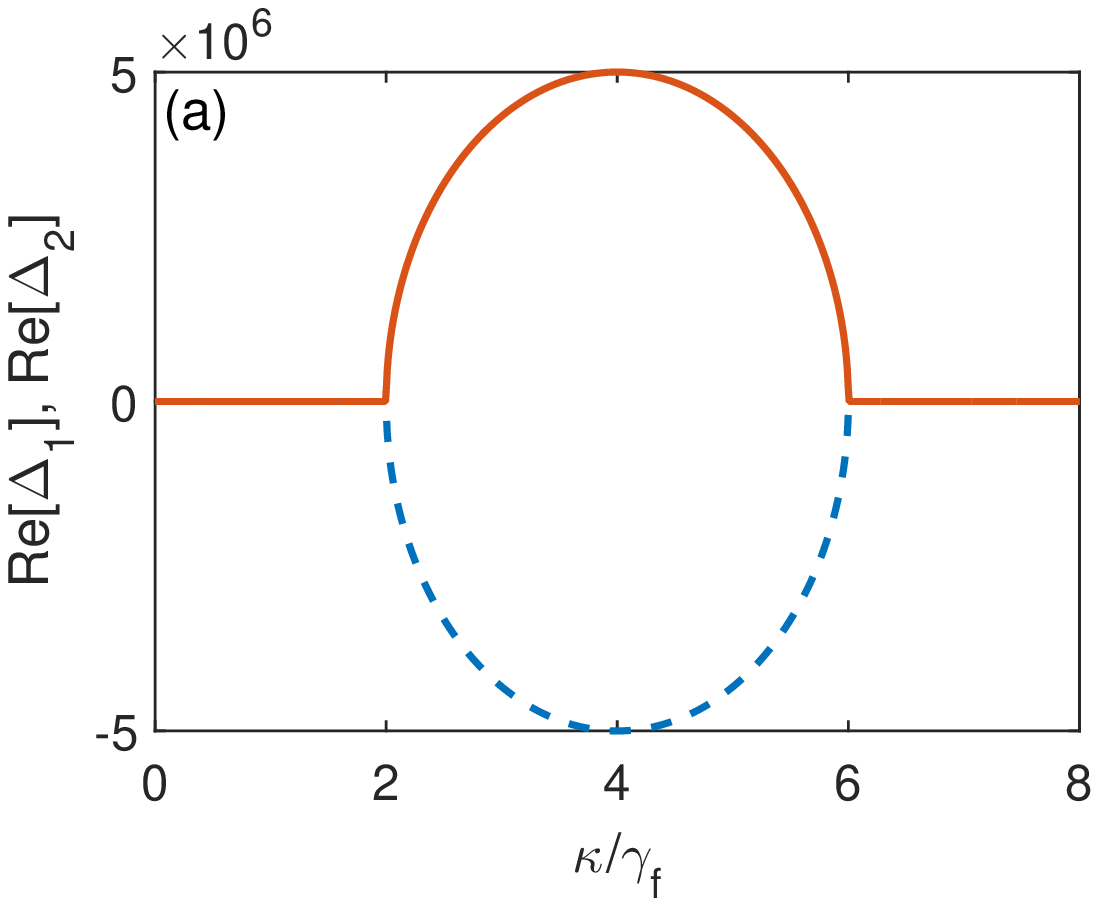}\includegraphics[width=6.5cm]{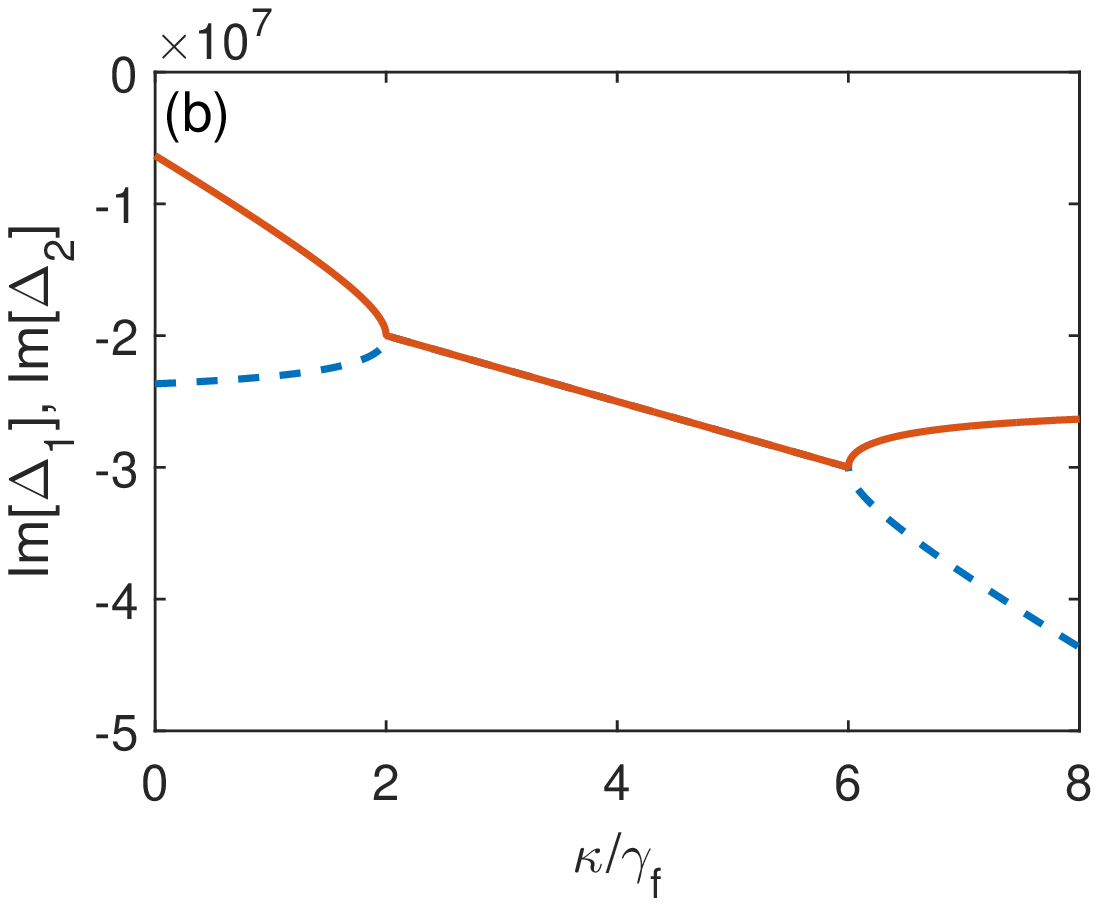}
\includegraphics[width=6.5cm]{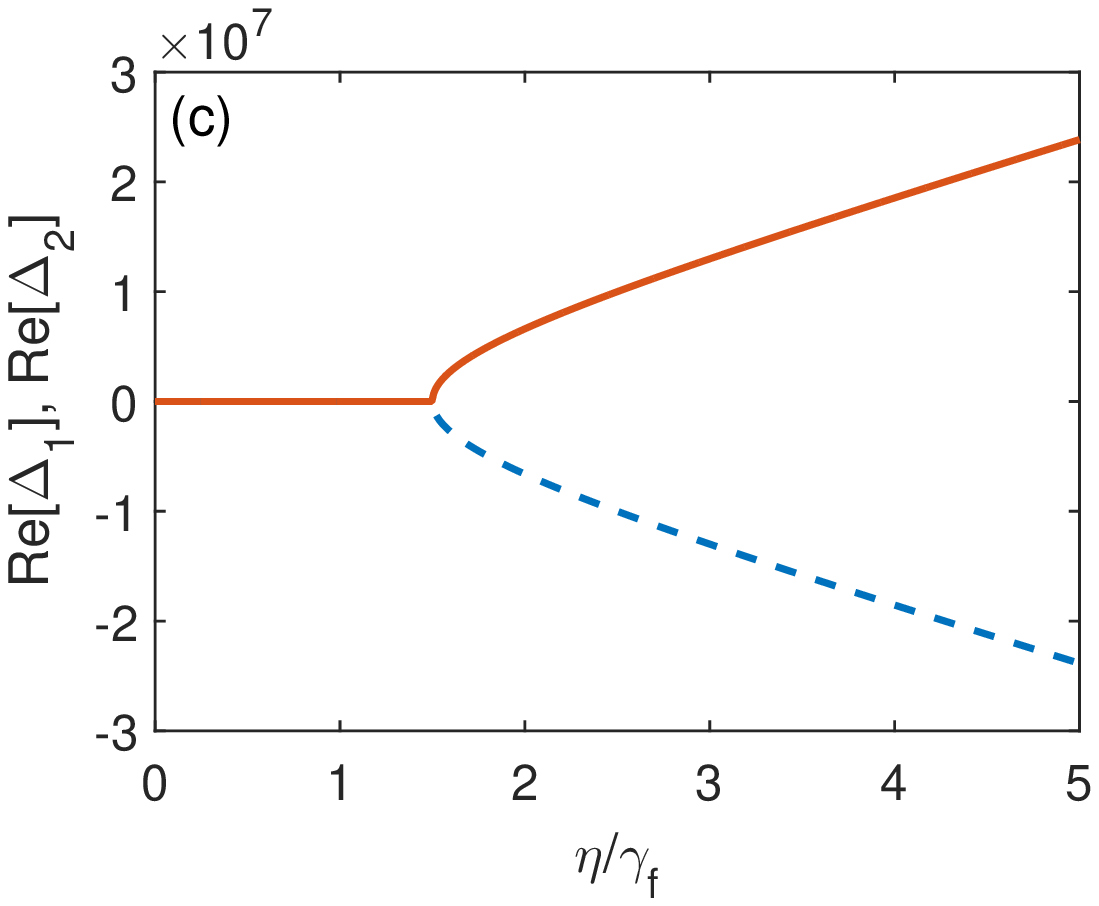} \includegraphics[width=6.5cm]{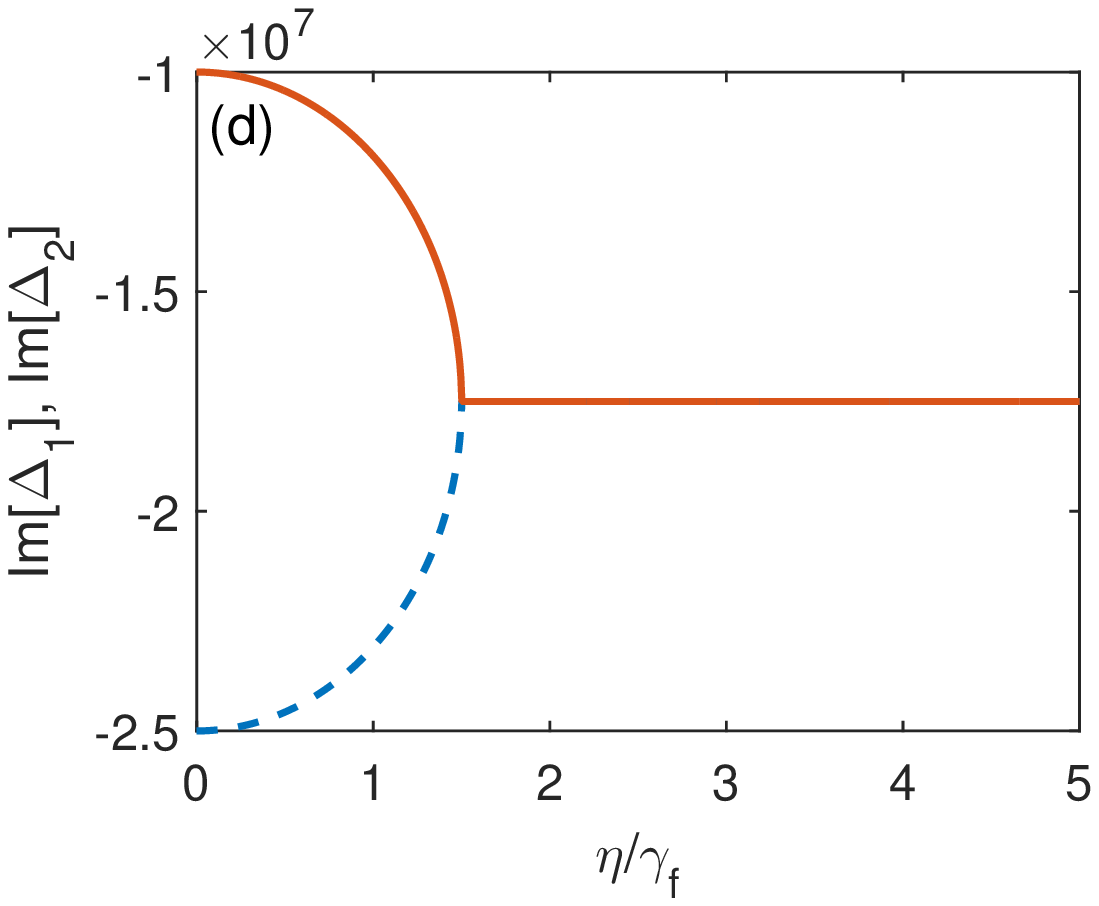}
\caption{(Color online) The real and the imaginary parts of $\Delta_{1}$ (red
solid) and $\Delta_{2}$ (blue dashed) in (a) and (b) are plotted
as the function of the cavity decay rate $\kappa$. Here, we assume
$\gamma_{e}=5\gamma_{f}$ and $\eta=4\gamma_{f}$. The real and the
imaginary parts of $\Delta_{1}$ (red solid) and $\Delta_{2}$ (blue
dashed) in (c) and (d) are plotted as the function of the coupling
strength $\eta$. Here, we assume $\gamma_{e}=5\gamma_{f}$ and $\kappa=\gamma_{f}$.}
\label{fig4}
\end{figure*}

\subsection{Vacuum induced transparency and Autler-Townes splitting}
\subsubsection{Resonance decomposition}
In analogy to ATS and EIT for a classically driven three-level system,
we further analyze physical mechanism of the dip and two peaks as shown in Fig.~\ref{fig3} when the three-level system is
coupled to the single-mode cavity field in vacuum case. Here, we study
the condition to discern VIT from vacuum induced ATS. To simplify discussions,
we only consider the case that the cavity field resonantly interacts
with the three-level system, i.e., $\delta=0$. Following the method
as in Refs.~\cite{Abi,sun14}, we first decompose the linear susceptibility
$\chi$ into two resonances. That is, using Eq.~\eqref{eq:susceptibility}
and Eq.~\eqref{eq:rho_ge_solution}, the susceptibility $\chi$ in
vacuum case can be decomposed as
\begin{eqnarray}
\chi & = & R_{1}(\Delta)+R_{2}(\Delta)\label{eq:chi_decomp}
\end{eqnarray}
with
\begin{eqnarray*}
R_{1}(\Delta) & \equiv & \frac{\beta}{\Delta_{1}-\Delta_{2}}\frac{\Delta_{1}+i(\gamma_{f}+\kappa)}{\Delta-\Delta_{1}},\\
R_{2}(\Delta) & \equiv & \frac{\beta}{\Delta_{1}-\Delta_{2}}\frac{-\Delta_{2}-i(\gamma_{f}+\kappa)}{\Delta-\Delta_{2}},
\end{eqnarray*}
with $\beta=\mu_{ge}/\epsilon_{0}$. Here, we note that the similar decomposition as in Eq.~(\ref{eq:chi_decomp}) can be obtained for
the expression of the susceptibility in Ref.~\cite{Vacuum-e}. That is, the resonance decomposition in Eq.~(\ref{eq:chi_decomp})
can be applied to analyze the formula in Ref.~\cite{Vacuum-e}.
Hereafter, we call $R_{1}(\Delta)$
and $R_{2}(\Delta)$ as the resonances. The parameters $\Delta_{1}$
and $\Delta_{2}$ are the complex roots of equation
\begin{equation}
\Delta^{2}+i(\gamma_{e}+\gamma_{f}+\kappa)\Delta-C=0,
\end{equation}
derived from Eq.~\eqref{eq:rho_ge_solution}. Here, the parameter
$C$ is given in Eq.~(\ref{eq:22}) with $\delta=0$. In this case,
$\Delta_{1}$ and $\Delta_{2}$ can be given by
\begin{eqnarray}
\Delta_{1} & = & \frac{1}{2}\left[-i(\gamma_{e}+\gamma_{f}+\kappa)+\sqrt{4\eta^{2}-\eta_{T}^{2}}\right],\\
\Delta_{2} & = & \frac{1}{2}\left[-i(\gamma_{e}+\gamma_{f}+\kappa)-\sqrt{4\eta^{2}-\eta_{T}^{2}}\right],
\end{eqnarray}
with $\eta_{T}=\left|\gamma_{f}+\kappa-\gamma_{e}\right|$. It is
clear that both $\Delta_{1}$ and $\Delta_{2}$ are pure imaginary
numbers when $4\eta^{2}-\eta_{T}^{2}<0$, but they are complex numbers
when $4\eta^{2}-\eta_{T}^{2}>0$. Below we further analyze how $\Delta_{1}$
and $\Delta_{2}$ change with the variations of different parameters.

In Fig.~\ref{fig4}(a) and Fig.~\ref{fig4}(b), the real and imaginary
parts of $\Delta_{1}$ and $\Delta_{2}$ are plotted as the function
of the cavity decay rate $\kappa$ for given parameters, e.g., $\eta=4\gamma_{f}$
and $\gamma_{e}=5\gamma_{f}$. From Fig.~\ref{fig4}(a) and Fig.~\ref{fig4}(b),
we find that $4\eta^{2}-\eta_{T}^{2}<0$ when $\kappa<2\gamma_{f}$
or $\kappa>6\gamma_{f}$. Both $\Delta_{1}$ and $\Delta_{2}$ are
pure imaginary numbers, and therefore their real parts are zero. However,
$4\eta^{2}-\eta_{T}^{2}>0$ when the decay rate $\kappa$ of the cavity
field is in the range $2\gamma_{f}<\kappa<6\gamma_{f}$, $\Delta_{1}$
and $\Delta_{2}$ become complex numbers. In this case their real
parts have different signs and the same amplitude, but their imaginary
parts have the same amplitude.

In Fig.~\ref{fig4}(c) and Fig.~\ref{fig4}(d), the real and the
imaginary parts of $\Delta_{1}$ and $\Delta_{2}$ are plotted as
the function of the coupling strength $\eta$ between the cavity field
and the three-level system for given parameters, e.g., $\kappa=\gamma_{f}$
and $\gamma_{e}=5\gamma_{f}$. This plot is similar to that in Ref.~\cite{Anisimov2008}. We find that
there are two different parameter regimes:
(i) the strong coupling regime when $\eta>\eta_{T}/2$ and (ii) the
weak coupling regime when $\eta<\eta_{T}/2$. In the weak coupling
regime, $\Delta_{1}$ and $\Delta_{2}$ are pure imaginary numbers
with different amplitudes, this means that the resonances $R_{1}(\Delta)$
and $R_{2}(\Delta)$ have the same frequency but different linewidths.
In the strong coupling regime, the real parts of $\Delta_{1}$ and
$\Delta_{2}$ have different signs and same amplitude. This means
that the resonances $R_{1}(\Delta)$ and $R_{2}(\Delta)$ have the
same linewidth and different frequencies. This property is quite similar
to that of PT-symmetric system~\cite{Cal,lan,jiang}. In the broken
PT-symmetric regime, the two coupled modes have different linewidths
while the frequencies are degenerate, this corresponds to the weak
coupling regime here for $\eta<\eta_{T}/2$. In the PT-symmetric regime,
the two modes have the same linewidth while the frequencies are different,
this corresponds to the strong coupling regime here for $\eta>\eta_{T}/2$.

\begin{figure*}
\includegraphics[width=13cm]{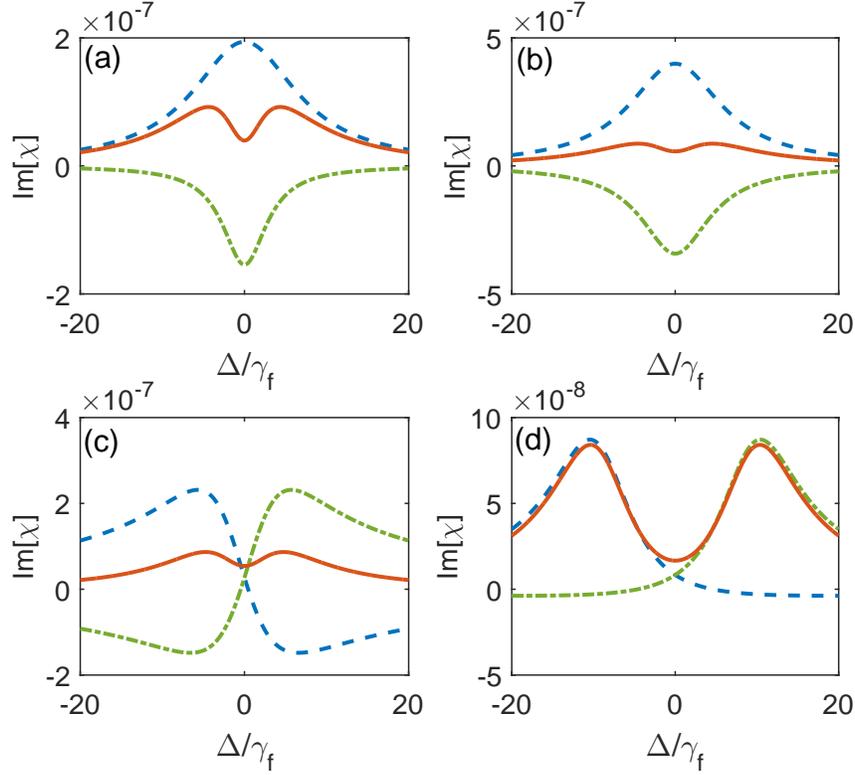}\\
 \caption{(Color online) Imaginary parts of the susceptibility $\mathrm{Im}[\chi]$
(red solid curve), the resonances $\mathrm{Im}[R_{1}(\Delta)]$ (blue
dashed curve) and $\mathrm{Im}[R_{2}(\Delta)]$ (green dash-dotted
curve) are plotted as the function of the detuning $\Delta$ between
the three-level system and the probe field for different parameters:
(a) $\gamma_{e}=10\gamma_{f}$, $\kappa=0$ and $\eta=3.9\gamma_{f}$
; (b) $\gamma_{e}=10\gamma_{f}$, $\kappa=\gamma_{f}$ and $\eta=3.9\gamma_{f}$;
(c) $\gamma_{e}=10\gamma_{f}$, $\kappa=\gamma_{f}$ and $\eta=4.1\gamma_{f}$;
(d) $\gamma_{e}=10\gamma_{f}$, $\kappa=\gamma_{f}$ and $\eta=10\gamma_{f}$.}
\label{fig5}
\end{figure*}

\subsubsection{Threshold condition for VIT and vacuum induced ATS}
To study the condition for realizing VIT and vacuum induced ATS, we
should first find the dip in the absorption spectrum. For convenience,
we call it as the dip condition. A dip in the absorption spectrum
means that the second derivative of $\mathrm{Im}[\chi]$ at $\Delta=0$
is positive. By solving $\partial^{2}\mathrm{Im}[\chi]/\partial\Delta^{2}>0$,
we obtain the dip condition
\begin{equation}
\eta>(\gamma_{f}+\kappa)\sqrt{\frac{\gamma_{f}+\kappa}{2(\gamma_{f}+\kappa)+\gamma_{e}}}.\label{eq:34}
\end{equation}
Besides the dip condition, we would like to find a critical value
to distinguish VIT and vacuum induced ATS. By applying $4\eta^{2}-\eta_{T}^{2}=0$,
we obtain the threshold
\begin{equation}
\eta=\frac{1}{2}\eta_{T}.\label{eq:35}
\end{equation}
It is clear that VIT and vacuum induced ATS can be distinguished from
this threshold condition. According to Ref.~\cite{sun14}, the vacuum
induced ATS should occur in the regime $\eta>\eta_{T}/2$, which is
called the strong-coupling regime in analogy to the strong driving
regime for the ATS in classically driven three-level system. In this
regime, two resonances have different frequencies and the same linewidth
as shown in Fig.~\ref{fig4}. While VIT should occur in the regime
$\eta<\eta_{T}/2$, which is called the weak-coupling regime in analogy
to the weak driving regime for EIT in classically driven three-level
system. In this regime, two resonances have different linewidths and
the same frequency as shown in Fig.~\ref{fig4}. Therefore, these
two coupling regimes can be distinguished by observing the shapes
of $\mathrm{Im}[R_{1}(\Delta)]$ and $\mathrm{Im}[R_{2}(\Delta)]$.
In vacuum induced ATS regime, both $\mathrm{Im}[R_{1}(\Delta)]$ and
$\mathrm{Im}[R_{2}(\Delta)]$ are positive Lorentzian shapes centered
at $\Delta=\sqrt{4\eta^{2}-\eta_{T}^{2}}$ and $\Delta=-\sqrt{4\eta^{2}-\eta_{T}^{2}}$,
respectively. While in VIT regime, $\mathrm{Im}[R_{1}(\Delta)]$ and
$\mathrm{Im}[R_{2}(\Delta)]$ have Lorentzian shapes with opposite
signs centered at $\Delta=0$.

Combining the dip condition in Eq.~(\ref{eq:34}) and the threshold
condition in Eq.~(\ref{eq:35}), we obtain the condition to realize
VIT
\begin{equation}
(\gamma_{f}+\kappa)\sqrt{\frac{\gamma_{f}+\kappa}{2(\gamma_{f}+\kappa)+\gamma_{e}}}<\eta<\frac{1}{2}\eta_{T},\label{eq:eit_condi}
\end{equation}
and that to realize vacuum induced ATS
\begin{equation}
\eta>\max\left\{ \frac{1}{2}\eta_{T},(\gamma_{f}+\kappa)\sqrt{\frac{\gamma_{f}+\kappa}{2(\gamma_{f}+\kappa)+\gamma_{e}}}\right\} ,\label{eq:ats_condi}
\end{equation}
for the coupling strength $\eta$. Note that, if VIT regime exists,
the relation $(\gamma_{f}+\kappa)\sqrt{(\gamma_{f}+\kappa)/[2(\gamma_{f}+\kappa)+\gamma_{e}]}<\eta_{T}/2$
should also be satisfied, which leads to
\begin{equation}
\gamma_{e}>2(\gamma_{f}+\kappa).\label{eq:eit_exist_condi}
\end{equation}
To simplify the notation, we introduce two parameters $\gamma_{R}$
and $\eta_{R}$ to interpret the conditions in Eqs.~\eqref{eq:eit_condi}-\eqref{eq:eit_exist_condi}.
Here, we define $\gamma_{R}=\gamma_{e}/(\gamma_{f}+\kappa)$ and $\eta_{R}=\eta/(\gamma_{f}+\kappa)$.
Using $\gamma_{R}$ and $\eta_{R}$, Eqs.~\eqref{eq:eit_condi}-\eqref{eq:eit_exist_condi}
can be rewritten as
\begin{eqnarray}
 &  & \eta_{d}<\eta_{R}<\eta_{c},\label{eq:crit_eit}\\
 &  & \gamma_{R}>2,
\end{eqnarray}
for VIT and
\begin{eqnarray}
 &  & \eta_{R}>\eta_{c},\gamma_{R}>2,\\
 &  & \eta_{R}>\eta_{d},\gamma_{R}<2,\label{eq:crit_ats}
\end{eqnarray}
for vacuum induced ATS with
\begin{eqnarray}
\eta_{d} & = & \frac{1}{\sqrt{2+\gamma_{R}}},\label{eq:eta_d}\\
\eta_{c} & = & \frac{1}{2}\left\vert 1-\gamma_{R}\right\vert .\label{eq:eta_c}
\end{eqnarray}
From above derivations, we conclude that the condition to realize
VIT and vacuum induced ATS is similar to that of EIT and ATS, but
there are differences. In both VIT and vacuum induced ATS, the dip
condition not only depends on the atomic decay rates $\gamma_{f}$
and $\gamma_{e}$, but also the decay rate $\kappa$ of the cavity,
which can be found from Eq.~(\ref{eq:34}). The condition for VIT is
$\gamma_{e}>2(\gamma_{f}+\kappa)$, however the condition for EIT
is $\gamma_{e}>2\gamma_{f}$, thus VIT is more difficult to be realized
in comparing with EIT.

In Fig.~\ref{fig5}, we show the imaginary part $\mathrm{Im}[\chi]$
of the susceptibility $\chi$ for absorption spectrum, the resonances
$\mathrm{Im}[R_{1}(\Delta)]$ and $\mathrm{Im}[R_{2}(\Delta)]$ for
several sets of possible values of the parameters. Compared to conventional
EIT, the cavity decay $\kappa$ here plays a crucial role in realizing
VIT. To highlight the effect of $\kappa$, we assume $\gamma_{f}$
is much smaller than $\gamma_{e}$ and $\eta$, this condition may
be satisfied in certain physical systems. According to the criteria in
Eqs.~\eqref{eq:crit_eit}-\eqref{eq:crit_ats}, we can verify that
Figs.~\ref{fig5}(a) and (b) are in VIT regime, while Figs.~\ref{fig5}(c)
and (d) are in vacuum induced ATS regime. Comparing Fig.~\ref{fig5}(a)
(where $\kappa=0$) with Fig.~\ref{fig5}(b) (where $\kappa=\gamma_{f}$),
we know that the cavity decay $\kappa$ has a negative effect in VIT
as we discussed before. In Figs.~\ref{fig5}(b) and (c), we show
the curves of $\mathrm{Im}[\chi]$ for $\eta=3.9\gamma_{f}$ and $\eta=4.1\gamma_{f}$,
but other parameters are the same. We can find the transition from
VIT in Fig.~\ref{fig5}(b) to vacuum induced ATS in Fig.~\ref{fig5}(c)
with a change of the coupling strength $\eta$. In Fig.~\ref{fig5}(d),
the coupling strength is further increased and two Lorentzian peak
appear. The curve for $\mathrm{Im}[\chi]$ is completely in vacuum
induced ATS regime. We mention that Akaike¡¯s information
criterion has been proposed as an objective test to discern the best model for
experimentally  obtained absorption  or  transmission  spectra~\cite{ATS-EIT} when the
 data are inconclusive. This criterion can also be applied to analyze the
 experimental data of VIT and vacuum induced ATS.

We note that our study here is very different from those in Refs.~\cite{ian10,H.Y.Zhang},
where they studied EIT and ATS using three energy levels chosen from
many energy levels of a two-level system dressed by a cavity field.
In their study, both the classical control and probe fields are applied
to the selected three-level system. The role of the cavity field in
Refs.~\cite{ian10,H.Y.Zhang} is to assist the two-level system realizing
a three-level system. These systems in Refs.~\cite{ian10,H.Y.Zhang}
are still used to study a conventional EIT and ATS. However, we here
study a three-level system, in which two upper energy levels are coupled
to a quantized single-mode cavity field, one and only one classical
probe field is applied to the system. The quantized cavity field acts as
a control field. Thus we study quantized-field-induced
quantum interference and frequency shift, we here call them as VIT
and vacuum induced ATS. Therefore, the cavity field in our study here
and in Refs.~\cite{ian10,H.Y.Zhang} has very different purpose.

\section{Photon resolved Aulter-Townes splitting}

In Sec.~IV, we mainly analyze the absorption of probe field by the three-level system when the quantized control
field is in vacuum at the steady state. This is realized by a zero temperature assumption of the environment. In this section,
we further study the absorption when the quantized control field contains the finite number of photons at the steady state.
This situation is different from the classical control field with very large number of photons for EIT and ATS, also different
from the quantized control field without photon for VIT and vacuum induced ATS. The finite photon number of quantized control
field might be realized by incoherent pumping or coherent pumping to the quantized control field.

\subsection{Incoherent pumping}

We assume that the incoherent pumping is realized by taking the environmental temperature $T$ into account. In this case,
the master equation in Eq.~(\ref{eq:master_equation}) is further modified to
\begin{eqnarray}
\dot{\rho} & = & \frac{1}{i\hbar}[H,\rho]+(n_{th}+1)\kappa(2a\rho a^{\dagger}-a^{\dagger}a\rho-\rho a^{\dagger}a)\nonumber \label{eq:master_equation_nth}\\
 &  & +n_{th}\kappa(2a^{\dagger}\rho a-aa^{\dagger}\rho-\rho aa^{\dagger}) \\
 &  & +\sum\limits _{i,j=e,f,g}^{i\geqslant j}\frac{\gamma_{ij}}{2}[2\sigma_{ji}\rho\sigma_{ij}-\sigma_{ii}\rho-\rho\sigma_{ii}]\nonumber
\end{eqnarray}
with the thermal photon $n_{th}=1/(e^{\hbar\omega/k_{b}T}-1)$
of the quantized control field.  Here, for simplicity and with loss of generality,
we have neglected effect of the temperature on the three-level system.

\begin{figure}
\includegraphics[width=8.5cm]{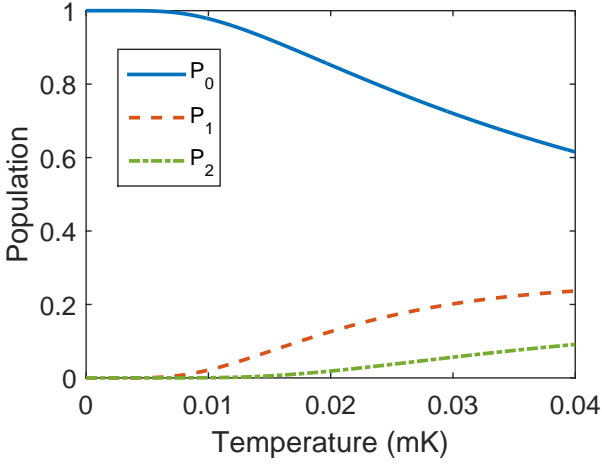} \caption{(Color online) The populations in the states $|0,g\rangle$, $|1,g\rangle$
and $|2,g\rangle$ as a function of the environmental temperature $T$
when the system reaches the steady state. Here, we assume that the
three-level system resonantly interacts with the quantized control field, i.e.,
$\delta=0$. $P_{0}$ (blue solid curve), $P_{1}$ (red dashed curve)
and $P_{2}$ (green dash-dotted curve) represent the population in
$|0,g\rangle$, $|1,g\rangle$ and $|2,g\rangle$, respectively. Here
we assume $\gamma_{e}=5\gamma_{f}$, $\kappa=0.2\gamma_{f}$ and $\eta=2\gamma_{f}$,
that is, we use $\gamma_{f}$ as units in our numerical calculations.}
\label{fig6}
\end{figure}

At the finite temperature, $n_{th}\neq0$, thus not all
population remains in the ground state $|0,g\rangle$ when the system
reaches steady state, other states, e.g., the states $|1,g\rangle$,
$|2,g\rangle$, $\cdots$, $|n,g\rangle$, may also be occupied. That
is, all possible states of the system might be involved in zero-order
solutions of all matrix elements of the density matrix. Therefore,
it is very difficult to obtain analytical solutions of $\chi$ at
the finite temperature. From Eq.~(\ref{eq:susceptibility}), we know
that the susceptibility $\chi$ is only related to states $|n,g\rangle$
and $|n,e\rangle$. To observe how the temperature affects the population
in different states, we numerically solve Eq.~(\ref{eq:master_equation_nth})
by truncating photon numbers to $60$. In Fig.~\ref{fig6}, we show
how the populations in, e.g., the states $|0,g\rangle$, $|1,g\rangle$
and $|2,g\rangle$, vary with the temperature $T$ for $\varepsilon=0$
when the system reaches the steady state under the resonant interaction between
the three-level system and the quantized control field. As shown
in Fig.~\ref{fig6}, when the environmental temperature is below
$10$ mk, almost all population is in the ground $|0,g\rangle$, the
population in other states is negligibly small. However, with the
increase of the temperature, the population in the state $|0,g\rangle$
is decreased, the populations in other states, e.g., $|1,g\rangle$
and $|2,g\rangle$, are increased.

\begin{figure}
\includegraphics[width=8.5cm]{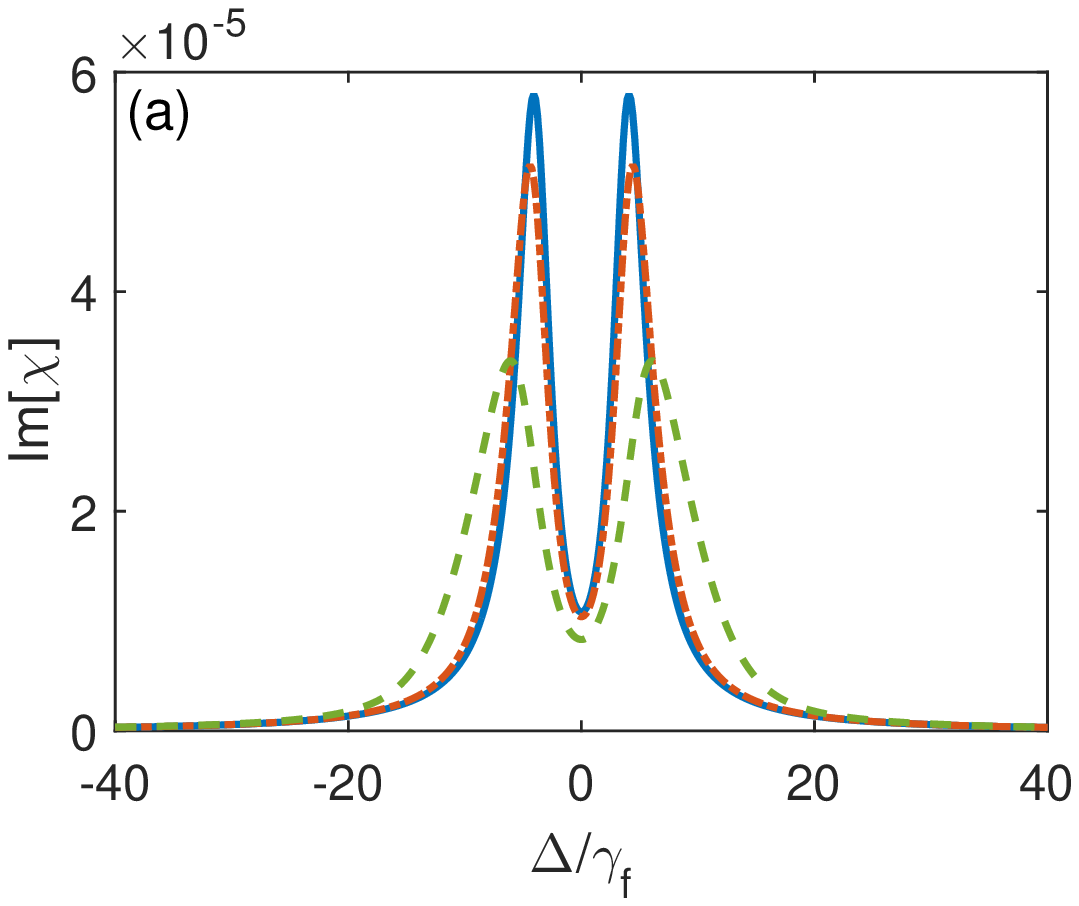} \includegraphics[width=8.5cm]{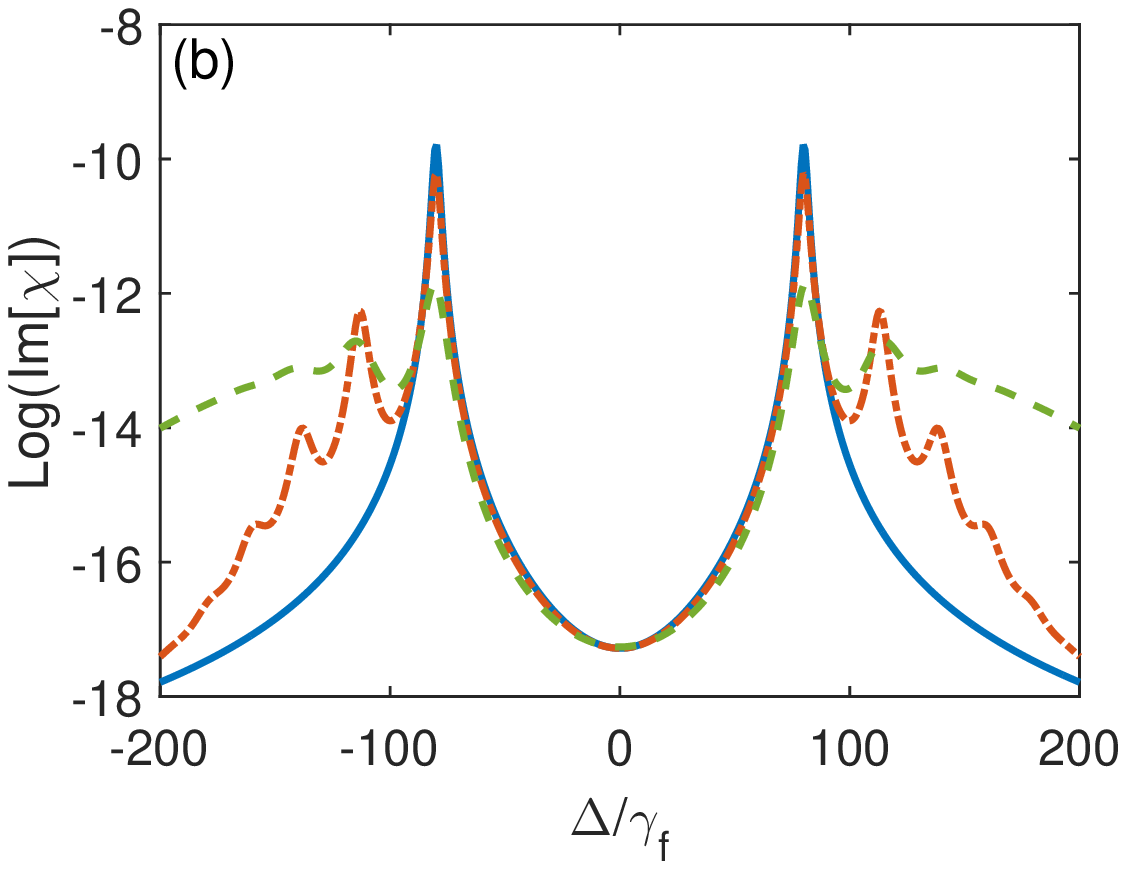}
\caption{The imaginary part $\text{Im}[\chi]$ of the susceptibility $\chi$
is plotted as the function of the detuning $\Delta$ between the three-level
system and the probe field when the three-level system resonantly
interacts with the quantized control field. (a) shows the absorption spectra
in the weak coupling regime, e.g., $\eta=4\gamma_{f}$. (b) shows
the absorption spectra in the strong coupling regime, e.g., $\eta=80\gamma_{f}$.
In each figure, we take three temperatures: zero temperature (blue
solid curve); $10$ mK (red dash-dotted curve), $80$ mK (green dashed
curve). Here, we assume $\gamma_{e}=5\gamma_{f}$, $\kappa=\gamma_{f}$.}
\label{fig7}
\end{figure}

Using master equation in Eq.~(\ref{eq:master_equation_nth}), $\mathrm{Im}[\chi]$
is plotted as a function of the detuning $\Delta$ in Fig.~\ref{fig7}
under the resonant interaction (i.e., $\delta=0$) between the three-level
system and the quantized control field. Fig.~\ref{fig7} shows $\mathrm{Im}[\chi]$
versus $\Delta$ when the three-level system and the quantized field
are either in weak or strong coupling regime for VIT or vacuum
induced ATS at the temperatures $T=0$ mK, $T=10$ mK and $T=80$
mK, respectively. At the zero temperature, the quantized control field is in vacuum when the whole system
reaches steady state, thus $\mathrm{Im}[\chi]$ is
only proportional to $\langle e,0|\rho|0,g\rangle$ as discussed in Sec.~IV. Under certain condition, two
peaks can appear in the absorption spectra as shown in Fig.~\ref{fig3} and Fig.~\ref{fig5}.
At the finite temperature, there are populations in the states $|n,g\rangle$
with $n>0$. Thus, $\mathrm{Im}[\chi]$ should include not only $\langle e,0|\rho|0,g\rangle$
for vacuum but also $\langle e,n|\rho|n,g\rangle$ with $n>0$ for
the finite photon number. In the weak coupling regime for VIT, the
distances between any two peaks of different photon numbers in the
same side of the spectrum are smaller than the linewidth of each peak,
thus we cannot observe photon number resolved peaks. However, in the
strong coupling regime for vacuum induced ATS, the distances between
any two peaks of different photon numbers in the same side of the
spectrum can be larger than the linewidth of each peak, thus photon
resolved peaks can be observed. For example, two peaks approximately
locate at $-\eta$ and $\eta$ in the spectrum for vacuum case corresponding
to $\langle e,0|\rho|0,g\rangle$, but two peaks locate at $-\eta\sqrt{2}$
and $\eta\sqrt{2}$ in the spectrum for single-photon case corresponding
to $\langle e,1|\rho|1,g\rangle$. When $(\sqrt{2}-1)\eta>\gamma_{e}$,
single-photon peak and vacuum peak can be resolved.

Fig.~\ref{fig7}(a) shows $\mathrm{Im}[\chi]$ versus $\Delta$ in
the weak coupling regime for VIT. When the temperature of the system is
not high, $\mathrm{Im}[\chi]$ for the finite photon number at finite temperature almost overlaps
with that for vacuum at zero temperature, but the heights of two peaks in absorption
spectrum are slightly reduced due to the effect of the temperature. This is because only single occupation plays
a role, other occupations are negligibly small. When the
temperature is further increased more than $10$ mK, as shown in Fig.~\ref{fig6},
the more states $|n,g\rangle$ ($n>0$) for higher energy levels are
involved. Although $\mathrm{Im}[\chi]$ is proportional to the summation
of all possible matrix elements $\langle e,n|\rho|n,g\rangle$ as
shown in Eq.~(\ref{eq:rho_ge}), the photon-number dependent
peaks in absorption spectrum are not resolved in the weak coupling regime. As a result, (i) the linewidths
of two peaks are broadened and the heights of two peaks are suppressed; (ii)
the position of each peak is shifted comparing with that at zero temperature,
because the peak of each component corresponding to $\langle e,n|\rho|n,g\rangle$
in $\mathrm{Im}[\chi]$ is slightly different, the peak position of
the summation for all these peaks is different from anyone of these peaks.
Such photon number dependent EIT might be used to realize photon number dependent
group delay for the probe field.
However, in the strong coupling case, as shown in Fig.~\ref{fig7}(b),
when the temperature of the system is zero, $\mathrm{Im}[\chi]$ exhibits
two peaks, we call them as vacuum induced ATS. When the temperature
of the system is further increased, e.g., $T=10$ mK, $\mathrm{Im}[\chi]$
exhibits even number of peaks, we call them as photon number resolved Aulter-Townes spectrum. But if the
temperature is further increased, the linewidth of each peak also
becomes larger, then many peaks for photon number resolved spectrum
will gradually become into two peaks.

\begin{figure}
\includegraphics[width=8.5cm]{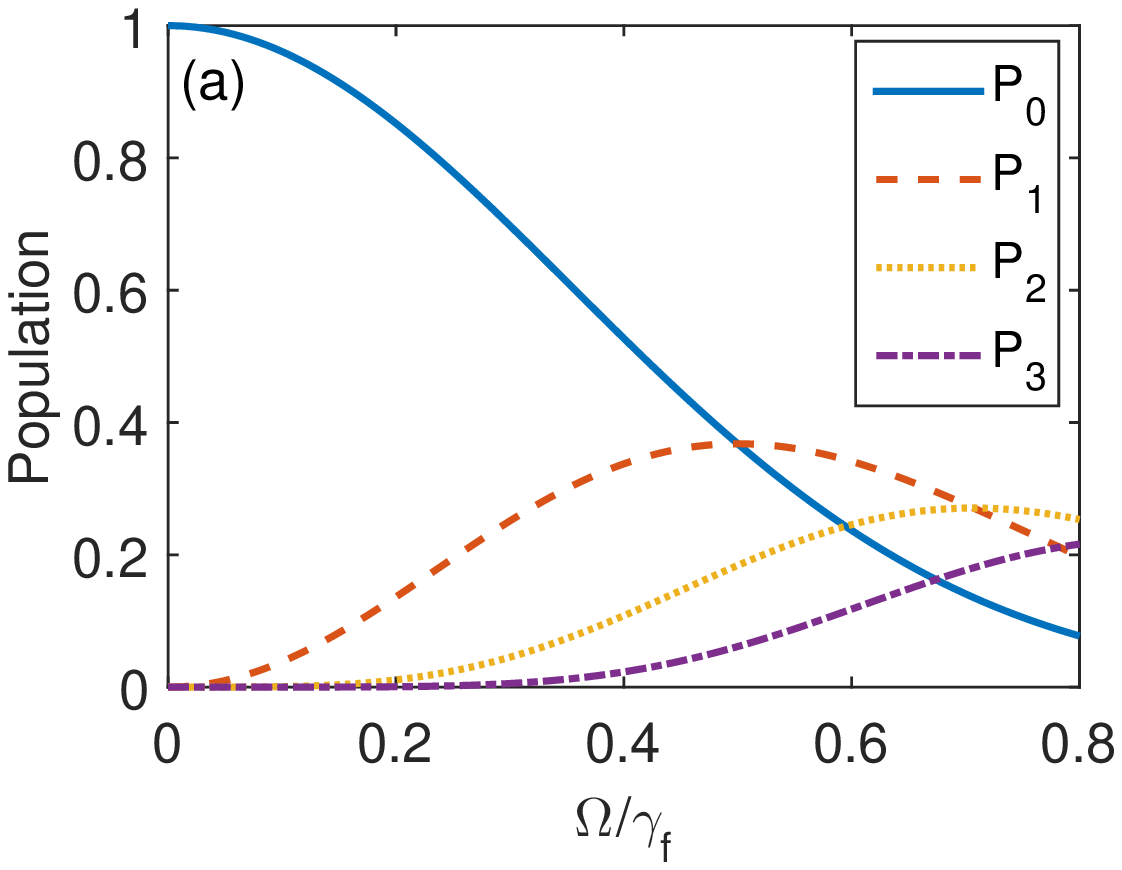}
\includegraphics[width=8.5cm]{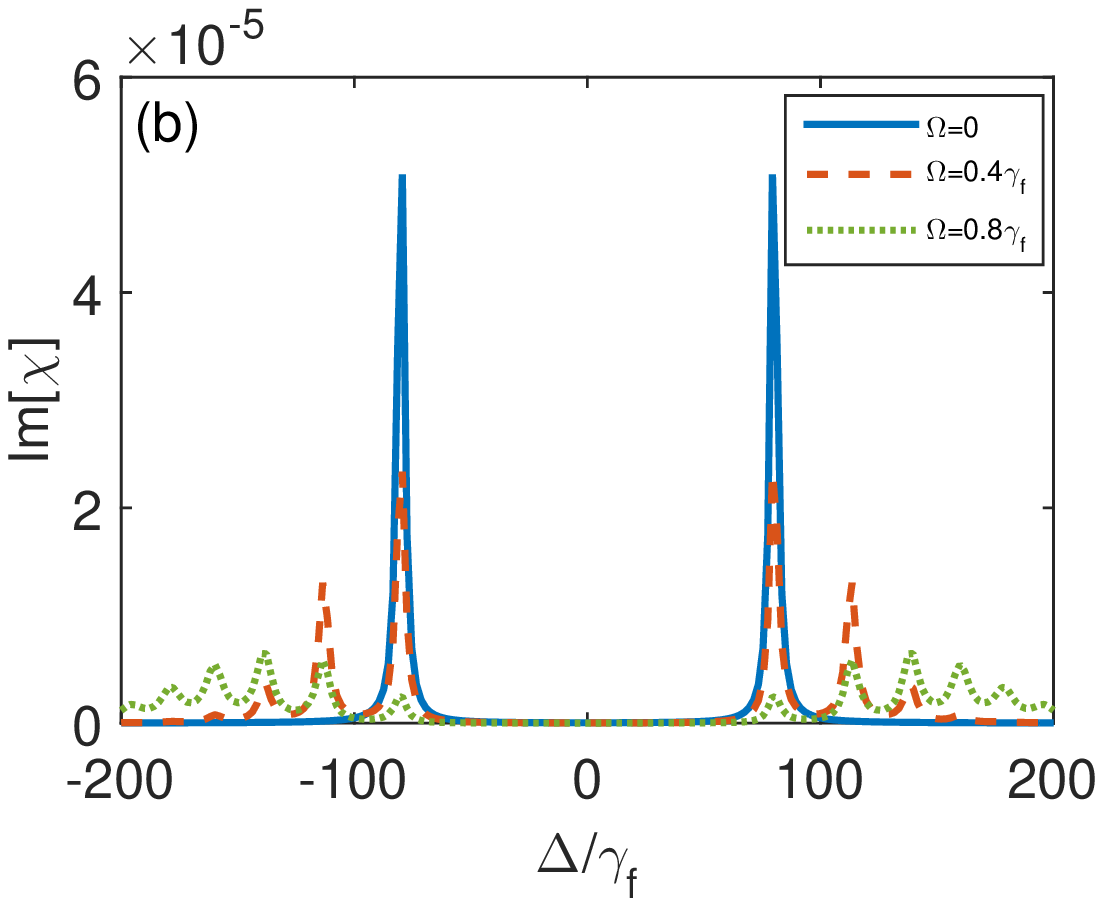}
\caption{(Color online) (a) The populations in the states $|0,g\rangle$, $|1,g\rangle$,
$|2,g\rangle$ and $|3,g\rangle$ as a function of the driving strength
$\Omega$ when the system reaches the steady state. Here, we assume
that the three-level system resonantly interacts with the quantized control field,
i.e., $\delta=0$. $P_{0}$ (blue solid curve), $P_{1}$ (red dashed
curve), $P_{2}$ (yellow dotted curve), $P_{3}$ (purple dash-dotted
curve) represent the populations in $|0,g\rangle$, $|1,g\rangle$,
$|2,g\rangle$ and $|3,g\rangle$, respectively. (b) The imaginary
part $\text{Im}[\chi]$ of the susceptibility $\chi$ versus the detuning
$\Delta$. Here we assume $\gamma_{e}=5\gamma_{f}$, $\kappa=\gamma_{f}$
and $\eta=80\gamma_{f}$, that is, we use $\gamma_{f}$ as units in
our numerical calculations. }
\label{fig8}
\end{figure}

\subsection{Coherent pumping}

We now consider another case that the quantized control field is pumped by a weak coherent
field. In this case, the Hamiltonian in Eq.~(\ref{ham_twopart}) is modified to
\begin{eqnarray}
H_{\text{Coh}} & = & \hbar\omega_{c}a^{\dagger}a+\hbar\omega_{e}\left\vert e\right\rangle \left\langle e\right\vert +\hbar\omega_{f}\left\vert f\right\rangle \left\langle f\right\vert \label{ham_driving}\\
 &  & +\hbar\eta(\left\vert e\right\rangle \left\langle f\right\vert a+\text{h.c.})\nonumber \\
 &  & +\hbar\varepsilon(\left\vert e\right\rangle \left\langle g\right\vert e^{-i\omega_{p}t}+\text{h.c.})+\hbar\Omega(a^{\dagger}e^{-i\omega_{d}t}+\text{h.c.}),\nonumber
\end{eqnarray}
where $\omega_{d}$ is the frequency of the pumping field and
$\Omega$ is the coupling strength between the pumping field and quantized control field.
Similar to the effect of thermal photons, a coherent pumping field
can also modify the occupations of photons in different states when the whole system reaches
the steady state. This will also result in photon number dependent absorption to the
probe field. To solely consider the pumping effect and without loss of generality, we assume that the whole system is at zero temperature. In this case, replacing $H$ in Eq.~(\ref{eq:master_equation}) by $H_{\text{Coh}}$ in Eq.~(\ref{ham_driving}),
we can numerically study the absorption spectrum by solving the
master equation. In our simulation, the photon number is truncated to $60$.

As shown in Fig.~\ref{fig8}(a), all the population remains in the ground $|0,g\rangle$
when the pumping field is not applied, i.e., $\Omega=0$. With the increase of the strength
$\Omega$ of the pumping field, the population occupation in the state $|0,g\rangle$ is decreased
while the populations in the states $|1,g\rangle$, $|2,g\rangle$ and $|3,g\rangle$ are increased.
In Fig.~\ref{fig8}(b), we show the imaginary part $\text{Im}[\chi]$
of the susceptibility $\chi$ versus the detuning $\Delta$ when the three-level system
and the quantized control field is in the strong coupling regime for vacuumed induced ATS. When
$\Omega=0$, as shown in the blue solid curve in Fig.~\ref{fig8}(b),
the absorption spectrum corresponds to vacuum induced ATS, which is discussed in Sec.~IV~B. When $\Omega=0.4\gamma_{f}$,
as shown in the red dashed curve in Fig.~\ref{fig8}b, the peaks located
at $\Delta=\pm\eta$, $\pm\sqrt{2}\eta$ and $\pm\sqrt{3}\eta$ emerge.
The heights of these peaks correspond to the populations in the states $|0,g\rangle$,
$|1,g\rangle$, $|2,g\rangle$ and $|3,g\rangle$ in Fig.~\ref{fig8}(a)
at $\Omega=0.4\gamma_{f}$. When $\Omega=0.8\gamma_{f}$, the population
in the state $|0,g\rangle$ is smaller than that in each of the states $|1,g\rangle$, $|2,g\rangle$ and
$|3,g\rangle$, and the corresponding peaks located at $\Delta=\pm\eta$
are also lower than others.

We find that the absorption spectrum of the probe field in coherent pumping is very similar to that
of the incoherent pumping when the coupling strength between the three-level system and the quantized control field is
in the parameter regime of VIT. That is, there might be two peaks and one dip in the absorption spectrum. With the increase of the strength of the pumping field, the heights of two peaks are reduced and the linewidths of two peaks are broadened.  When the
strength of the pumping field is further increased, the quantized control field approaches to the classical one, then the spectrum approaches to that of EIT. There is a difference for the absorption spectra between the coherent pumping and the incoherent pumping. When the quantized control field is coherently pumped, the steady-state population in the state $\left\vert n,g\right\rangle $ ($n\neq0$) can exceed that in the state $\left\vert 0,g\right\rangle $ when the strength $\Omega$ of the pumping field is large enough. If the coupling strength between the three-level system and the quantized control field is in the parameter regime of vacuum induced ATS, as shown in Fig.~\ref{fig8}(b) for the strong pumping case (the green dotted curve), then we have photon number resolved spectrum. The peaks located in $\Delta=\pm\eta$ correspond to the transition from $\left\vert 0,g\right\rangle $ to $\left\vert 0,e\right\rangle $. The peaks located in $\Delta=\pm\sqrt{n}\eta$  correspond to the transition from $\left\vert n,g\right\rangle $ to $\left\vert n,e\right\rangle $ ($n\neq0$). From Fig.~\ref{fig8}(a), we can find that the steady-state population in the state $\left\vert 0,g\right\rangle $ is smaller than that in the state $\left\vert n,g\right\rangle $ ($n\neq0$) when $\Omega=0.8\gamma_{f}$. This makes that the heights of peaks located in $\Delta=\pm\eta$ are lower than those of peaks located in $\Delta=\pm\sqrt{n}\eta$. However, when the quantized control field is incoherently pumped, the steady-state population in the state $\left\vert 0,g\right\rangle $ is always larger than those in other states $\left\vert n,g\right\rangle $ ($n\neq0$).

\section{Discussions on possible applications to superconducting quantum circuits}

Our study here can in principle
be applied to any $\Lambda$-type three-level system, coupled to a quantized single-mode cavity field and detected by
a weak probe field, as schematically shown in Fig.~\ref{fig1}. For example, a three-level
atomic system, coupled to a quantized field~\cite{Vacuum-e,Boller}, is good for demonstrating VIT
because of smaller decay rate of the first excited state and weaker coupling
strength between the three-level system and quantized control field,
but atomic systems might not be easy to demonstrate vacuum induced
ATS because the coupling strength between the three-level system and
the quantized control field is not very strong.

Let us now explore another possibility to demonstrate these phenomena
using superconducting circuit QED system~\cite{circuitQED-R},
which is extensively studied for quantum information processing and
quantum optics on superconducting chip~\cite{you-today,clarke08,you-nature}.
For concreteness of discussions, we assume that the three-level system
with $\Lambda$-type transitions is constructed by a superconducting
flux qubit circuit~\cite{Orlando,Mooij,YXLiu2005} when the magnetic
flux bias deviates from the optimal point. Recently, such a three-level
flux qubit circuit is used to demonstrate the correlated microwave
lasing~\cite{Peng} by coupling it to two modes of a coplanar waveguide
resonator. Using experimentally accessible parameters in Ref.~\cite{Peng},
e.g., decay rates $\gamma_{e}=2\pi\times7.5$ MHz and $\gamma_{f}=2\pi\times3.25$
MHz of three-level system, and decay rate $\kappa=2\pi\times0.63$
MHz for one of modes, and coupling strength $\eta=2\pi\times36$ MHz
between this mode and three-level system, we can obtain $\gamma_{R}$=1.93,
$\eta_{R}$=9.28, $\eta_{d}$=0.50 and $\eta_{c}$=0.47. Evaluating
the parameters $\gamma_{R}$, $\eta_{R}$, $\eta_{d}$ and $\eta_{c}$
with the conditions in Eqs.~\eqref{eq:crit_eit}-\eqref{eq:crit_ats},
we find that the vacuum induced ATS can be demonstrated using this
set of experimental parameters. Moreover, it is possible to demonstrate
photon number resolved ATS in this system. We also find that the value
of $\gamma_{R}$, with parameters in Ref.~\cite{Peng}, is very close
to $2$. Thus, if the experimental parameters can be further optimized
so that $\gamma_{R}>2$, and also a proper coupling strength $\eta$
can be chosen, then VIT can be realized in such superconducting three-level
system coupled to a single mode microwave field.

Although we only discuss possible realization in a three-level superconducting
flux qubit circuit which is coupled to a microwave cavity field, the
study here can also be applied to phase~\cite{ATS-3,ATS-4,Martinis}
and other superconducting quantum circuits, which possess $\Lambda$-type
transitions. We mention that the inversion symmetry of the potential
energy for superconducting flux~\cite{Orlando,Mooij,YXLiu2005},
transmon~\cite{transmon1,transmon} and Xmon~\cite{Xmon} qubit
circuits is well defined, thus the transition from the ground state
to the second excited state is forbidden at the optimal point. They
have ladder-type transitions and no $\Lambda$-type transitions at
the optimal point. How VIT and vacuum induced ATS occur in such a
ladder-type three-level system is still under study.

\section{Conclusions}

In conclusion, we have studied the absorption spectrum of a probe field by
a $\Lambda$-type three-level system, in which two upper energy levels
are coupled to a quantized single-mode control field.
If the quantized control field is replaced by a classical control field,
then the system is usually studied for EIT and ATS. We find that there are
similarity and difference in the absorption spectra for the classical and quantized control field.
(1) If the quantized control field is in vacuum, then the vacuum induced absorption spectrum is very similar to EIT and
or ATS spectrum. That is, there is a transparency windows formed by two peaks and a dip in the absorption spectrum.
In the parameter regime of the weak coupling between the quantized control field and three-level
system, VIT might occur. Similar to EIT, the distance between two peaks in the absorption spectrum for VIT is smaller than two times of the coupling strength.
Moreover, we find that VIT is more difficult to be realized than EIT when the cavity leakage is included.
That is, the cavity decay plays a negative role in the realization of VIT.
In the strong coupling regime, vacuum induced ATS occur. Similar to ATS, the distance between two peaks in the absorption spectrum for vacuum induced ATS
is two times of the coupling strength.
(2) If the quantized control field contains finite number of photons, then in the weak coupling regime, absorption spectrum is also
similar to that of EIT. Only difference is that the heights of two peaks will be suppressed and the linewidths of two peaks are broadened with the
increase of the photon number. When photon number is further increased, the quantized control field approaches classical one,
then the spectrum approaches to that of EIT. In particular, in the strong coupling regime, we find the so-called photon number resolved ATS,
which is very different from ATS. There are even number of peaks in the absorption spectrum. We mention that there were experiments on photon number resolved spectrum~\cite{P1,P2} when a superconducting qubit (a two-level system) is strongly coupled to a quantized field.  There, two-level system is dispersively coupled to the quantized field, the photon number is observed by virtue of the ac Stark shifted qubit frequency~\cite{P1,P2}. Here, a three-level system is resonantly coupled to the quantized field, the photon number is observed by virtue of the resonant absorption spectrum.

Comparing with studies for VIT~\cite{Vacuum-t,Vacuum-e}, we here give the threshold condition to discern VIT from vacuum induced ATS
in such a system. The realization of VIT requires that the coupling
strength between the three-level system and the quantized control field is
smaller than a critical value, which depends on the damping rates
of the three-level system and the quantized control field. However,
the realization of the vacuum induced ATS requires that the coupling
strength between the three-level system and the quantized control field is larger
than the critical value. We also show that the parameter changing from VIT to vacuum induced
ATS is very similar to that from broken $PT$ symmetry to $PT$ symmetry.  Furthermore, we
studied the photon number dependent spectrum, in particular, we show a photon number resolved ATS in the parameter
regime of the strong coupling between the quantized control field and three-level
system

We also explore possible experiments using natural atomic systems or superconducting
quantum circuits. We find that three-level natural atomic systems
might be a good candidate to demonstrate VIT because of smaller decay
rate of the first excited state and weaker coupling
strength  between three-level system and quantized control field,
but atomic systems might not be easy to demonstrate vacuum induced
ATS because the coupling strength between the three-level system and
the quantized control field is not very strong. However, vacuum induced ATS is
easy to be demonstrated in the superconducting quantum circuits because
the coupling strength between three-level superconducting
qubit circuit and  the quantized control field can be very strong, but VIT may not be easy to be demonstrated.
Thus, to show VIT in superconducting quantum system, the decay rates
of two excited states of three-level system should be further optimized.

In summary, we study the quantized field controlled absorption spectrum
in a three-level system. In particular, we give a threshold to discern
VIT from vacuum induced ATS. We also find photon number resolved ATS, which is very different from ATS and vacuum induced ATS.
We hope that our study can motivate more experiments to realize photon
control for weak probe field at single-atom and single-photon level.

\begin{acknowledgments}

Y.X.L. acknowledges the support of the National Basic Research Program
of China Grant No. 2014CB921401 and the National Natural Science Foundation
of China under Grant No. 91321208. H. I. acknowledges the support
of FDCT Macau under grant 013/2013/A1, University of Macau under grant
MRG022/IH/2013/FST and MYRG2014-00052-FST, and the National Natural
Science Foundation of China under Grant No.~11404415.
\end{acknowledgments}

\appendix

\section{Calculation of equation of motion of matrix elements}

In dressed state picture, the operators $\sigma_{ij}$ and $a^{\dagger}$
with their hermitian conjugate $\sigma_{ji}$ and $a$ can be given
by
\begin{eqnarray}
\sigma_{ge} & = & \cos\theta_{0}\left\vert G\right\rangle \left\langle u\right\vert +\sin\theta_{0}\left\vert G\right\rangle \left\langle v\right\vert ,\\
\sigma_{fe} & = & \cos\theta_{0}\left\vert 0,f\right\rangle \left\langle u\right\vert +\sin\theta_{0}\left\vert 0,f\right\rangle \left\langle v\right\vert ,\\
\sigma_{gf} & = & \left\vert G\right\rangle \left\langle 0,f\right\vert ,\\
a^{\dagger} & = & \left\vert g^{^{\prime}}\right\rangle \left\langle G\right\vert -\sin\theta_{0}\left\vert u\right\rangle \left\langle 0,f\right\vert +\cos\theta_{0}\left\vert v_{0}\right\rangle \left\langle 0,f\right\vert ,\\
a^{\dagger}a & = & \left\vert g^{^{\prime}}\right\rangle \left\langle g^{^{\prime}}\right\vert +\sin^{2}\theta_{0}\left\vert u\right\rangle \left\langle u\right\vert +\cos^{2}\theta_{0}\left\vert v\right\rangle \left\langle v\right\vert \nonumber \\
 &  & -\sin\theta_{0}\cos\theta_{0}\left\vert u\right\rangle \left\langle v\right\vert -\sin\theta_{0}\cos\theta_{0}\left\vert v\right\rangle \left\langle u\right\vert .
\end{eqnarray}
Note that $\left\vert 0,f\right\rangle $ isn't belong to any subspace,
if we replace the term that contain $\left\vert 0,f\right\rangle $
by zero, the result remain to be the same.

Since $\rho_{ge}$ is the superposition of $\rho_{Gu}$ and $\rho_{Gv}$,
we substitute the above equations and the dressed Hamiltonian H$_{c}$
in Eq.~(\ref{eq:4}) and H$_{p}$ in Eq.~(\ref{eq:hp_dressed})
into Eq.~(\ref{eq:master_equation}), then we can list all terms
of the equations of motion for $\rho_{Gu}$ and $\rho_{Gv}.$
\begin{eqnarray*}
\dot{\rho}_{Gv} & = & \left\langle G\right\vert \dot{\rho}\left\vert v\right\rangle \\
\dot{\rho}_{Gu} & = & \left\langle G\right\vert \dot{\rho}\left\vert u\right\rangle
\end{eqnarray*}
Because, we have
\begin{eqnarray*}
\frac{1}{i\hbar}\left\langle G\right\vert [H,\rho]\left\vert v\right\rangle  & = & \frac{1}{i\hbar}[-E_{+0}\rho_{Gv}+\hbar\varepsilon\cos\theta_{0}e^{i\omega_{p}t}\rho_{uv},\\
 &  & +(\rho_{vv}-\rho_{GG})\hbar\varepsilon\sin\theta_{0}e^{i\omega_{p}t}],\\
\frac{1}{i\hbar}\left\langle G\right\vert [H,\rho]\left\vert u\right\rangle  & = & \frac{1}{i\hbar}[-E_{-0}\rho_{Gu}+\hbar\varepsilon\sin\theta_{0}e^{i\omega_{p}t}\rho_{vu},\\
 &  & +(\rho_{uu}-\rho_{GG})\hbar\varepsilon\cos\theta_{0}e^{i\omega_{p}t}],\\
\left\langle G\right\vert D_{ge}(\rho)\left\vert v\right\rangle  & = & -\cos\theta_{0}\sin\theta_{0}\rho_{Gu}-\sin^{2}\theta_{0}\rho_{Gv},\\
\left\langle G\right\vert D_{fe}(\rho)\left\vert v\right\rangle  & = & -\cos\theta_{0}\sin\theta_{0}\rho_{Gu}-\sin^{2}\theta_{0}\rho_{Gv},\\
\left\langle G\right\vert D_{ee}(\rho)\left\vert v\right\rangle  & = & -\cos\theta_{0}\sin\theta_{0}\rho_{Gu}-\sin^{2}\theta_{0}\rho_{Gv},\\
\left\langle G\right\vert D_{gf}(\rho)\left\vert v\right\rangle  & = & \cos\theta_{0}\sin\theta_{0}\rho_{Gu}-\cos^{2}\theta_{0}\rho_{Gv},\\
\left\langle G\right\vert D_{ff}(\rho)\left\vert v\right\rangle  & = & \cos\theta_{0}\sin\theta_{0}\rho_{Gu}-\cos^{2}\theta_{0}\rho_{Gv},\\
\left\langle G\right\vert D_{aa^{\dagger}}(\rho)\left\vert v\right\rangle  & = & \cos\theta_{0}\sin\theta_{0}\rho_{Gu}-\cos^{2}\theta_{0}\rho_{Gv},\\
\left\langle G\right\vert D_{ge}(\rho)\left\vert u\right\rangle  & = & -\cos^{2}\theta_{0}\rho_{Gu}-\sin\theta_{0}\cos\theta_{0}\rho_{Gv},\\
\left\langle G\right\vert D_{fe}(\rho)\left\vert u\right\rangle  & = & -\cos^{2}\theta_{0}\rho_{Gu}-\sin\theta_{0}\cos\theta_{0}\rho_{Gv},\\
\left\langle G\right\vert D_{ee}(\rho)\left\vert u\right\rangle  & = & -\cos^{2}\theta_{0}\rho_{Gu}-\sin\theta_{0}\cos\theta_{0}\rho_{Gv},\\
\left\langle G\right\vert D_{gf}(\rho)\left\vert u\right\rangle  & = & -\sin^{2}\theta_{0}\rho_{Gu}+\sin\theta_{0}\cos\theta_{0}\rho_{Gv},\\
\left\langle G\right\vert D_{ff}(\rho)\left\vert u\right\rangle  & = & -\sin^{2}\theta_{0}\rho_{Gu}+\sin\theta_{0}\cos\theta_{0}\rho_{Gv},\\
\left\langle G\right\vert D_{aa^{\dagger}}(\rho)\left\vert u\right\rangle  & = & -\sin^{2}\theta_{0}\rho_{Gu}+\sin\theta_{0}\cos\theta_{0}\rho_{Gv},
\end{eqnarray*}
where we define $D_{ij}(\rho)=2\sigma_{ij}\rho\sigma_{ji}-\sigma_{jj}\rho-\rho\sigma_{jj}$
($i,j=g,f,e,a,a^{\dagger}$). Thus, we obtain the equations of motion
for $\rho_{Gu}$ and $\rho_{Gv}$ as shown in Eqs.~(\ref{eq:rho_gv})
and (\ref{eq:rho_gu}). % as below
%\begin{eqnarray}
%\dot{\rho}_{Gv} & = & \frac{1}{i\hbar}[-E_{+,0}\rho_{Gv}+(\rho_{vv}-\rho_{GG})\hbar\varepsilon\sin\theta_{0}e^{i\omega_{p}t}\notag\\
% &  & +\hbar\varepsilon\cos\theta_{0}e^{i\omega_{p}t}\rho_{uv}]\\
% &  & +\Gamma_{\mathrm{c}}\rho_{Gu}+\Gamma_{Gv}\rho_{Gv},\notag
%\end{eqnarray}
%\begin{eqnarray}
%\dot{\rho}_{Gu} & = & \frac{1}{i\hbar}[-E_{-,0}\rho_{Gu}+(\rho_{uu}-\rho_{GG})\hbar\varepsilon\cos\theta_{0}e^{i\omega_{p}t}\notag\\
% &  & +\hbar\varepsilon\sin\theta_{0}e^{i\omega_{p}t}\rho_{vu}]\\
% &  & +\Gamma_{\mathrm{c}}\rho_{Gv}+\Gamma_{Gu}\rho_{Gu},\notag
%\end{eqnarray}
%where we define the dressed relaxation rates $\Gamma_{Gv}$ and $\Gamma_{Gu}$
%as
%\begin{eqnarray*}
%\Gamma_{Gv} & = & -\gamma_{e}\sin^{2}\theta_{0}-(\gamma_{f}+\kappa)\cos^{2}\theta_{0},\\
%\Gamma_{Gu} & = & -\gamma_{e}\cos^{2}\theta_{0}-(\gamma_{f}+\kappa)\sin^{2}\theta_{0},
%\end{eqnarray*}
%and the crossing term relaxation
%\[
%\Gamma_{\mathrm{c}}=(-\gamma_{e}+\gamma_{f}+\kappa)\sin\theta_{0}\cos\theta_{0}.
%\]
%The calculations on the equations of motion for matrix elements with
%at most one photon case can be completed using the same method as
%for vacuum case.

\end{document}